\documentclass{jfm}
\usepackage{graphicx}
\usepackage{epstopdf, epsfig}
\usepackage{amsmath}
\usepackage{amsfonts}
\usepackage{mathrsfs}
\usepackage{amssymb}
\usepackage{xcolor}
\usepackage{multicol}
\usepackage{placeins}
\usepackage{bm}
\usepackage{upgreek}

\makeatletter
\renewcommand*\env@matrix[1][\arraystretch]{%
  \edef\arraystretch{#1}%
  \hskip -\arraycolsep
  \let\@ifnextchar\new@ifnextchar
  \array{*\c@MaxMatrixCols c}}
\makeatother

\shorttitle{Self-similar mechanisms in wall turbulence}
\shortauthor{U. Karban et al.}

\title{Self-similar mechanisms in wall turbulence studied using resolvent analysis}

\author{U. Karban\aff{1,2}
  \corresp{\email{ukarban@metu.edu.tr}}, E. Martini\aff{1}, A.V.G. Cavalieri\aff{3}, L. Lesshafft\aff{4}, P. Jordan\aff{1}}

\affiliation{
\aff{1}{D\'{e}partement Fluides, Thermique, Combustion, Institut Pprime, CNRS-University of Poitiers-ENSMA, France}
\aff{2}{Department of Aerospace Engineering, Middle East Technical University, Ankara 06800, Turkey}
\aff{3}{Instituto Tecnol\'{o}gico de Aeron\'{a}utica, S\~{a}o Jos\'{e} dos Campos/SP, Brazil}
\aff{4}{Laboratoire d’Hydrodynamique, CNRS / \'{E}cole polytechnique, Institut Polytechnique de Paris, Palaiseau, France}
}

\begin{document}

\maketitle
\begin{abstract}
Self-similarity of wall-attached coherent structures in a turbulent channel at $Re_\tau=543$ is explored by means of resolvent analysis. In this modelling framework, coherent structures are understood to arise as a response of the linearised mean-flow operator to generalised, frequency-dependent Reynolds stresses, considered to act as an endogenous forcing. We assess the self-similarity of both the wall-attached flow structures and the associated forcing. The former are educed from direct numerical simulation data by finding the flow field correlated with the wall shear, whereas the latter is identified using a frequency space version of Extended Proper Orthogonal Decomposition (Bor\'{e}e, J. 2003 Extended proper orthogonal decomposition: a tool to analyse correlated events in turbulent flows. \emph{Experiments in fluids} 35 (2), 188-192). The forcing structures identified are compared to those obtained using the resolvent-based estimation introduced by \citet{towne_jfm_2020} (Towne, A., Lozano-Dur\'{a}n, A. \& Yang, X. 2020 Resolvent-based estimation of space-time flow statistics. \emph{Journal of Fluid Mechanics} 883, A17). The analysis reveals self-similarity of both wall-attached structures---in quantitative agreement with Townsend's hypothesis of self-similar attached eddies---and the underlying forcing, at least in certain components.

\end{abstract}

\section{Introduction} \label{sec:intro}

The study of coherent structures in turbulent flow dates back to the early 1950s. Since the first observations of these organised motions, for instance by \citet{townsend_mp_1951,townsend_jfm_1976}, \citet{mollochristensen_jam_1967}, and \citet{crow_jfm_1971}, a substantial body of work has been dedicated to developing an understanding of how they work and what dynamical role they play. Reviews for wall-bounded flow have been provided by \citet{robinson_arm_1991} and \citet{jimenez_arm_2012}. 

Coherent structures can be seen in flow visualisations or simulations of boundary layers, where they are manifest in the form of streaks \citep{wu_jfm_2009,eitel_pof_2015,jodai_jfm_2016} or hairpin-like structures \citep{adrian_pof_2007}. They may be analysed quantitatively in terms of their statistics and compared with frequency space models \citep{sen_jot_2007,baltzer_ctr_2010,jordan_arm_2013,
lesshafft_prf_2019,cavalieri_amr_2019,abreu_jfm_2020,
morra_jfm_2021,nogueira_jfm_2021}. A recent overview of the different structures observed in wall-bounded turbulent flows, the focus of this study, can be found in \cite{lee_pof_2019}.

An early modelling idea, where wall-bounded flows are concerned, is that of the attached-eddy hypothesis (AEH), introduced by Townsend \citep{townsend_mp_1951,townsend_jfm_1976} and further developed by \citet{perry_jfm_1982}. Underlying this hypothesis is the idea that eddies in the logarithmic region of wall-bounded turbulent flows extend to the wall. This implies that their characteristic dimensions scale with distance to the wall, which implies, in turn, the existence of a self-similar organisation. A recent review of the literature on the AEH can be found in \citet{marusic_arm_2019}.

The original AEH considers coherent structures in the logarithmic region of boundary layers \citep{townsend_jfm_1976} at high-Reynolds number ($Re$), and predicts an $\alpha^{-1}$ decay in the 1-D energy spectrum, where $\alpha$ is the streamwise wavenumber. The $\alpha^{-1}$ decay was observed in wall-bounded turbulent flows at friction Reynolds number $Re_\tau\sim 100,000$ \citep{perry_jfm_1986,chandran_jfm_2017}, suggesting the existence of attached eddies that dominate the kinetic energy in the logarithmic region. The high Reynolds number of those studies makes it difficult to conclude on the validity of the AEH in flows simulated numerically or studied at laboratory scale with low or moderate Reynolds number. {\cite{cheng_jfm_2020} showed that wall-attached structures account for a significant portion of the flow energy even at low-$Re$ flows, ranging from $Re_\tau=180$ to 1000. They extracted the structures that move together and reach up to the wall, and observed geometric self-similarity among these structures.} \cite{chandran_prf_2020} investigated the 2-D spectra of the streamwise velocity component, where a transition from $\alpha^{-1/2}$ to $\alpha^{-1}$ scaling was observed as moving from {$Re_\tau=2400$ to 26000}. \citet{davidson_pof_2006,davidson_jfm_2006} showed that the $\alpha^{-1}$ decay is not observed at lower Reynolds number, and they attributed this to the limited range of scales of attached eddy that can exist at low $Re$, whence their smaller contribution to the energy spectrum. In view of this they suggested using the second-order structure function as a better indicator of attached eddies. \cite{agostini_prf_2017} used structure functions to show that attached-eddy behaviour can be observed at wall-normal distance, $y^+$, ranging from 80 to 2000 in channel flow at $Re_\tau=4200$. Their study demonstrated that choosing the correct method to identify attached eddies, as statistical entities, is crucial in finding evidence of AEH in turbulent flows that are attainable at lab-scale or via numerical simulation. 

There exist many techniques for the detection of coherent structures. One may use two-point measurements {\citep{tomkins_jfm_2003,monty_jfm_2007,baars_jfm_2017}}, conditional sampling {\citep{hussain_jfm_1986,hwang_jfm_2016_2,hwang_jfm_2020_2,cheng_jfm_2020}} or modal decomposition techniques such as proper orthogonal decomposition (POD) \citep{lumley_1967}, dynamic mode decomposition (DMD) \citep{schmid_jfm_2010}, or spectral proper orthogonal decomposition (SPOD) \citep{lumley_jfm_1970,picard_ijhff_2000,towne_jfm_2018,
schmidt_jfm_2018}. A review of these decomposition techniques, except SPOD, can be found in \citet{taira_aiaa_2017}. It has been demonstrated how  POD can be used to educe attached eddies in pipe flows at $Re_\tau=1300$ \citep{hellstrom_jfm_2016} and at $Re_\tau=685$ \citep{hellstrom_rsta_2017}. Those studies showed how leading POD modes are self-similar and scale with distance to the wall. This is in line with Townsend's original hypothesis which states that attached eddies correspond to the most energetic motions of the boundary layer \citep{townsend_jfm_1976}. 


Coherent structures obtained through the aforementioned decomposition techniques do not satisfy the Navier-Stokes equations. An attached eddy model based, for instance, on similarity analyses such as discussed above, provides a kinematic description only. A number of studies have shown, using dynamical descriptions, how the energy containing motions at different scales in turbulent wall-bounded flows may be maintained through a self-sustaining mechanism (see \cite{cossu_rsta_2017} for a review on the subject). Those studies rely on overdamped large-eddy simulations of the filtered Navier-Stokes equations \citep{hwang_prl_2010,hwang_jfm_2015}, and they show that the self-sustaining cycle exhibits self-similarity. We here use standard DNS coupled with the resolvent framework to investigate such self-similar dynamics for the wall-attached structures in the flow. 

Resolvent analysis provides a dynamical framework to study coherent structures and the non-linear interactions that drive them \citep{mckeon_jfm_2010,hwang_jfm_2010,schmidt_jfm_2018,
lesshafft_prf_2019,cavalieri_amr_2019}. The approach involves linearising the Navier-Stokes equations about the mean flow and retaining the non-linear terms as an inhomogeneous forcing of the linearised system. The problem can then be cast in an input/output (forcing/response) form, with forcing and response connected by the resolvent operator. Optimal forcing-response mode pairs can be identified from the resolvent operator, revealing the most important linear growth mechanisms in the flow. In many flows, the leading response mode is found to closely match coherent structures educed from time-resolved flow data \citep{mckeon_jfm_2010,lesshafft_prf_2019}. It has been shown in the literature (\cite{hwang_jfm_2010}, {\cite{moarref_jfm_2013}}, \cite{mckeon_jfm_2017} and {\cite{sharma_ptrsa_2017}} among others) that, for wall-bounded flows, the optimal response mode exhibits a self-similar structure reminiscent of attached eddies. However, the dynamics of coherent structures are ultimately determined by the details of the non-linear interactions at work in the flow, and these must be considered in order to understand how forcing/response modes of the resolvent operator combine to produce the observed behaviour  \citep{zare_jfm_2017,rosenberg_prf_2019,pickering_aiaa_2019,
martini_jfm_2020,morra_jfm_2021,nogueira_jfm_2021,symon_jfm_2021}.  Consideration of the resolvent operator alone can only provide a limited understanding of the dynamics of attached eddies in wall-bounded flows. It is this that motivates the study we undertake, where a data-driven approach is elaborated to explore the self-similarity of wall-attached coherent structures \textit{and} the non-linear interactions that drive them. {It is important to understand if the self-similarity seen in the flow structures is imposed only by the resolvent operator or if a self-similar forcing also plays a role. The former scenario implies that nonlinear terms, although not self-similar, are filtered by the self-similar resolvent operator such that the response is self-similar. The latter scenario, on the other hand, indicates a dynamic self-similarity, which can be useful for modeling wall-shear related phenomena in turbulent flows.}   


Our objective is to identify forcing structures that exist in the data and are associated with observed wall-attached coherent structures.
One approach that can be used to do this has been proposed by \citet{towne_jfm_2020}, where a resolvent-based estimation (RBE) of forcing statistics was achieved by combining flow measurements with the resolvent operator. \citet{martini_jfm_2020} extended that work by proposing an optimal estimation approach by which the space-time forcing field can be computed from a limited set of flow measurements; the estimator requires a model for the forcing statistics. RBE is recovered when a white model forcing is assumed, showing that this is an underlying assumption of the RBE method.


In this work we adopt an alternative approach, in which the extended proper orthogonal decomposition (EPOD) of \citet{boree_eif_2003}, which we cast in frequency space, is applied in the resolvent framework so as to identify forcing structures that are associated with an observed coherent structure. This method was first discussed in \cite{towne_aiaa_2015}. We refer to this tailored implementation of EPOD as Resolvent-based Extended Spectral Proper Orthogonal Decomposition (RESPOD). Related methods are the `observable inferred decomposition' (OID) \citep{schlegel_jfm_2012} and approaches based on linear stochastic estimation, such as in \citet{kerherve_jfm_2012}. In each of these cases, linear mappings are identified between a selected observable and some other field considered to drive that observable (the flow structures in a jet responsible for sound radiation, for example). But linear mappings so identified are not grounded in any rigorous, dynamic framework. The advantage of the RESPOD method is that a dynamic relation is granted within the resolvent framework. Empirically observed coherent structures are used to identify the forcing structures that exist in unsteady data and that drive the coherent structures via the resolvent operator. The dynamic relationship between forcing and response is identified such that it be consistent with the linearised Navier-Stokes equations. We use this approach to find coherent structures that are correlated to the wall shear, i.e., that are wall-attached, and the associated forcing. 

A related effort is the work of \cite{skouloudis_prf_2021}, who superpose resolvent modes with various wavenumbers, as a representation of attached eddies, so as to recover the Reynolds shear stress of channels. This is referred to as a quasi-linear approximation (QLA), which may be seen as a dynamic model of attached eddies based on the linearised Navier-Stokes operator. For simplicity, the superposition is restricted to zero streamwise wavenumber, and, for most cases, to zero frequency. As discussed in the cited paper, even with restricted wavenumbers and frequencies, such a superposition is non-unique, as there are several combinations of forcing modes that lead to the same overall Reynolds shear stress. While the approach of \cite{skouloudis_prf_2021} leads to correct Reynolds-number trends, a quantitative comparison with simulation data reveals differences. Information on non-linear terms driving flow responses may be built into dynamic attached-eddy models for more accurate predictions. We anticipate that self-similarity may be identified in both forcing and response modes with the techniques we aim to use in the present study. Use of the educed self-similar forcing in QLA is expected to lead to structures that better match fluctuations in wall-bounded turbulence.

As we will discuss (\S \ref{sec:method}), RESPOD is related to the RBE method of \citet{towne_jfm_2020}. RBE provides the `observable' forcing which has the minimal-norm required to generate the observed coherent structure, and it eliminates all `silent' components of forcing, i.e. those components that, while present in the data do not directly drive the response. We will show that RESPOD finds, for a given coherent structure, all the correlated forcing, including the silent components. These silent components, though redundant in terms of the direct driving of coherent structures through the resolvent operator, provide additional information regarding the forcing mechanisms at play, and, considered together with the `non-silent', driving components, provide a more complete picture of the forcing structures that actually exist in the unsteady data and are correlated with the observed coherent structures. We thus obtain a more complete description of attached eddies and the scale interactions by which they are driven.



This paper is organised as follows: in \S\ref{sec:method}, the RBE and RESPOD methods are revisited. The characteristics of the two methods are discussed via implementation on a toy model. The methods are then applied in a turbulent channel flow problem in \S\ref{sec:channel}. A DNS database with friction Reynolds number $Re_\tau=543$ is used. The methodology to trace attached eddies in the turbulent channels and to compute the associated forcing is presented with the results. Further discussions are provided in \S\ref{sec:conc}.
 
\section{Identifying the forcing associated with optimal response} \label{sec:method}
\subsection{Resolvent-based estimation} \label{subsec:rbe}
We consider the Navier-Stokes (N-S) equations,
\begin{align}
\mathbf{M}\partial_t\mathbf{q}(\mathbf{x},t)=\mathcal{N}\left(\mathbf{q}(\mathbf{x},t)\right),
\end{align}
where $\mathbf{q}=[\rho\,u\,v\,w\,p]^\top$ is the state vector, $\mathcal{N}$ denotes the nonlinear N-S operator and the matrix $\mathbf{M}$ is zero for the continuity equation and identity matrix for the rest in incompressible flows, or identity matrix in compressible flows. Discretisation in space and linearisation around the mean, $\overline{\mathbf{q}}(\mathbf{x})$, yields
\begin{align} \label{eq:forceintime}
\mathbf{M}\partial_t\mathbf{q}^{\prime}(\mathbf{x},t)-\mathbf{A}(\mathbf{x})\mathbf{q}^{\prime}(\mathbf{x},t)=\mathbf{f}(\mathbf{x},t),
\end{align}
where $\mathbf{A}(\mathbf{x})=\partial_q\mathcal{N}|_{\overline{\mathbf{q}}}$ is the linear operator obtained from the Jacobian of $\mathcal{N}$ and $\mathbf{f}(\mathbf{x},t)$ denotes all the remaining nonlinear terms, interpreted as a forcing term in the above equation. The equations are linearised using primitive variables \citep{karban_jfm_2020}. In the resolvent framework, \eqref{eq:forceintime} is Fourier transformed and rearranged to obtain 
\begin{align} \label{eq:forceinfreq}
\hat{\mathbf{q}}(\mathbf{x},\omega)=\mathbf{R}(\mathbf{x},\omega)\hat{\mathbf{f}}(\mathbf{x},\omega),
\end{align}
where $\mathbf{x}=[x\,y\,z]$ is the space vector, $\omega$ is the angular frequency, the hat indicates a Fourier transformed quantity and $\mathbf{R}(\mathbf{x},\omega)=(-i\omega\mathbf{M}-\mathbf{A}(\mathbf{x}))^{-1}$ is the resolvent operator. In what follows, notation showing dependence on $\mathbf{x}$ and $\omega$ will be dropped for brevity.

The cross-spectral densities of the response and forcing can also be related via the resolvent operator,
\begin{align} \label{eq:forceinfreqcsd}
{\mathbf{S}}=\mathbf{R}{\mathbf{P}}\mathbf{R}^H,
\end{align}
where $\mathbf{S}=E\{\hat{\mathbf{q}}\hat{\mathbf{q}}^H\}$ and $\mathbf{P}=E\{\hat{\mathbf{f}}\hat{\mathbf{f}}^H\}$ denote the response and forcing cross-spectral density (CSD) matrices, respectively, with $E\{\cdot\}$ denoting the expectation operator obtained by averaging different realizations, and the superscript $H$ denoting the conjugate transpose, or Hermitian.  

The resolvent operator can be modified in order to explore the relationship between subsets of the forcing and response. For instance, limited measurements at a selection of points, $\mathbf{\hat{y}}$, may be considered and related to forcing,
\begin{align} 
\hat{\mathbf{y}}&=\mathbf{C}\hat{\mathbf{q}},\\
\hat{\mathbf{y}}&=\tilde{\mathbf{R}}\hat{\mathbf{f}},\label{eq:modifres}
\end{align}
where $\mathbf{C}$ denotes the measurement matrix and $\tilde{\mathbf{R}}\triangleq\mathbf{C}\mathbf{R}$ is the modified resolvent operator that connects forcing to those measurements.
 We will first investigate the case where $\mathbf{C}=\mathbf{I}$ and $\mathbf{R}$ is full-rank, i.e. $\tilde{\mathbf{R}}$ is also full-rank, before extending our analysis to cases where $\tilde{\mathbf{R}}$ may be singular. Note that equations \eqref{eq:forceintime} and \eqref{eq:forceinfreq} are exact and provide a bijective relation between the forcing and the response when the resolvent operator is full-rank, and an injective relation when $\mathbf{R}$ is singular. 


We employ SPOD to identify coherent structures in the flow. This requires definition of an inner product,
\begin{align} \label{eq:inner}
\langle \mathbf{a},\mathbf{b}\rangle = \int_\Omega \mathbf{b}^H\boldsymbol{W}\mathbf{a}\,d\mathbf{x},
\end{align}
where $\boldsymbol{W}$ results in an energy norm. Equation \eqref{eq:inner} can be written in discrete form as
\begin{align} \label{eq:innerdis}
\langle \mathbf{a},\mathbf{b}\rangle = \mathbf{b}^H\mathbf{W}\mathbf{a},
\end{align}
where $\mathbf{W}$ now accounts both for the energy norm and the numerical quadrature weights. The SPOD modes, which are orthogonal with respect to the inner product defined by \eqref{eq:innerdis}, are obtained by solving the eigenvalue problem,
\begin{align} \label{eq:spoddis}
\mathbf{S}\mathbf{W}\bm{\psi} = \uplambda\bm{\psi},
\end{align}
where $\bm{\psi}$ and $\uplambda$ denote the eigenvector (SPOD mode) and the eigenvalue (SPOD mode mean square value). The CSD matrix, $\mathbf{S}$, can be built from SPOD modes as,
\begin{align} \label{eq:responseeig}
\mathbf{S} = \sum_n \uplambda_n\bm{\psi}_n\bm{\psi}_n^H,
\end{align}
where the subscript $n$ denotes SPOD mode number.

For non-singular $\mathbf{R}$, the forcing mode that generates a given SPOD mode, $\bm{\psi}$, can be obtained by inverting the resolvent equation,
\begin{align} \label{eq:frcinv}
\bm{\phi}=\mathbf{L}\bm{\psi},
\end{align}
where $\mathbf{L}\triangleq(-i\omega\mathbf{M}-\mathbf{A})$ and $\mathbf{R}=\mathbf{L}^{-1}$. In the case of $\tilde{\mathbf{R}}$, which is singular, direct inversion is not possible, and the forcing mode associated with the SPOD mode, $\bm{\psi}$ can then be obtained by means of a pseudo-inverse of $\tilde{\mathbf{R}}$,
\begin{align} \label{eq:frcpseudo}
\bm{\phi}=\tilde{\mathbf{R}}^+\bm{\psi}.
\end{align}
The mode $\bm{\phi}$ in \eqref{eq:frcpseudo} can be understood as the minimal-norm forcing that creates the response, $\bm{\psi}$, through the resolvent operator, in what is called resolvent-based estimation (RBE) \citep{towne_jfm_2020,martini_jfm_2020}. The forcing structures that are correlated with the response but are `silent', i.e., have no flow response, are not present in the forcing modes estimated using RBE. This might be a desired property for the kinematic modelling of forcing, as whatever is included in the estimated mode is ensured to contribute to the response. However, from a dynamical modelling point of view, elimination of the correlated-but-silent parts of the forcing may hide important dynamic traits of the interaction mechanisms that underpin the forcing structures. Since the silent components are correlated to the observable forces, they are likely generated by the same mechanisms. Thus, their study can facilitate identification of these mechanisms. To identify such structures, we consider, in what follows, a data-driven approach that takes into account the forcing and response structures that are actually contained in the data.

\subsection{Resolvent-based extended spectral proper orthogonal decomposition} \label{subsec:espod}
The approach we present is adapted from the EPOD of \citet{boree_eif_2003}, whose goal was to identify correlations between an observed POD mode and some other target field. In \cite{hoarau_prf_2006}, the method was used in spectral domain. We revisit EPOD also in frequency space, setting the target event as the forcing defined by equation \eqref{eq:forceinfreq}. This specific construction of EPOD was first discussed in \cite{towne_aiaa_2015}. We refer here to this implementation as `resolvent-based, extended spectral proper orthogonal decomposition' (RESPOD).

The SPOD modes defined by \eqref{eq:spoddis} are orthogonal with respect to the norm defined by \eqref{eq:innerdis}, i.e.,
\begin{align} \label{eq:spodort}
\langle \bm{\psi}_n,\bm{\psi}_p\rangle = \delta_{np},
\end{align} 
where $\bm{\psi}_n$ is the $n$th SPOD mode. The SPOD modes provide a complete basis that can be used to expand any realisation of the state vector as,
\begin{align} \label{eq:spodexpan}
\hat{\mathbf{q}} = \sum_n a_n\bm{\psi}_n,
\end{align}
where the projection coefficient, $a_n$, associated with the $n$th eigenmode is obtained by the projection,
\begin{align}
a_n = \langle \hat{\mathbf{q}},\bm{\psi}_n\rangle.
\end{align}
The orthonormality of the SPOD basis, defined by equation \eqref{eq:spodort}, imposes that the projection coefficients satisfy,
\begin{align} \label{eq:coeffrand}
E\{a_n a_p^H\} = \uplambda_n\delta_{np}.
\end{align}
\textcolor{black}{This can be seen as follows,
\begin{align} \label{eq:coeffrand2}
E\{a_n a_p^H\} &= E\{\langle \hat{\mathbf{q}},\bm{\psi}_n\rangle\langle \hat{\mathbf{q}},\bm{\psi}_n\rangle^H\}, \nonumber\\
&= E\{\bm{\psi}_n^H\mathbf{W}\hat{\mathbf{q}}\hat{\mathbf{q}}^H\mathbf{W}^H\bm{\psi}_p\}  \textrm{ (using \eqref{eq:innerdis})}, \nonumber\\
&= \bm{\psi}_n^H\mathbf{W}E\{\hat{\mathbf{q}}\hat{\mathbf{q}}^H\}\mathbf{W}^H\bm{\psi}_p,\nonumber\\
&= \bm{\psi}_n^H\mathbf{W}\sum_k \uplambda_k\bm{\psi}_k\bm{\psi}_k^H\mathbf{W}^H\bm{\psi}_p  \textrm{ (using \eqref{eq:responseeig})}, \nonumber\\
&= \uplambda_n\delta_{np}.
\end{align}
}

Here, the expectation operator corresponds to an ensemble average of Fourier realisations. \citet{boree_eif_2003} used equations \eqref{eq:spodexpan} and \eqref{eq:coeffrand} to show that,
\begin{align}
\begin{split}
E\{\hat{\mathbf{q}}\,a_p^H\}&=E\left\lbrace\left(\sum_n a_n\bm{\psi}_n\right)a_p^H\right\rbrace, \\
&=\sum_nE\{a_na_p^H\}\bm{\psi}_n, \\
&=\uplambda_p\bm{\psi}_p,
\end{split}
\end{align}

which provides an alternative way to compute $\bm{\psi}_p$ as,
\begin{align} \label{eq:psialt}
\bm{\psi}_p=\frac{E\{\hat{\mathbf{q}}\,a_p^H\}}{\uplambda_p},
\end{align}
assuming that $a_p$ is known.
\textcolor{black}{
Equation \eqref{eq:psialt} corresponds to the snapshot approach, shown by \citet{towne_jfm_2018} to provide a less costly alternative to eigendecomposition of \eqref{eq:spoddis}. Note that for this, one needs to calculate the projection coefficients beforehand. Using the decomposition given in \eqref{eq:spodexpan}, we can show that,
\begin{align} \label{eq:qinner}
\langle \hat{\mathbf{q}},\hat{\mathbf{q}}\rangle = \sum_n a_n a_n^H.
\end{align}
For a given $\hat{\mathbf{Q}}$ defined as the set of realisations of $\hat{\mathbf{q}}$, \eqref{eq:qinner} can be written as 
\begin{align} \label{eq:Qinner}
\langle \hat{\mathbf{Q}},\hat{\mathbf{Q}}\rangle = \sum_n \mathbf{a}_n {\mathbf{a}_n}^H,
\end{align}
where $\mathbf{a}_n\triangleq\langle\hat{\mathbf{Q}},\bm{\psi}_n\rangle$ denotes the projection coefficient vector. Note that \eqref{eq:coeffrand} holds also for $\mathbf{a}_n$ implying the orthogonality of the projection coefficient vectors that correspond to different SPOD modes. Given this orthogonality, rewriting \eqref{eq:Qinner} by normalising $\mathbf{a}_n$ with $\uplambda_n^{-1/2}$ as,
\begin{align} \label{eq:Qinnereig}
\langle \hat{\mathbf{Q}},\hat{\mathbf{Q}}\rangle = \sum_n (\mathbf{a}_n\uplambda_n^{-1/2})\uplambda_n (\mathbf{a}_n^H\uplambda_n^{-1/2}),
\end{align}
we obtain the eigendecomposition of $\langle \hat{\mathbf{Q}},\hat{\mathbf{Q}}\rangle$. Computing the eigendecomposition given in \eqref{eq:Qinnereig}, one can obtain the projection coefficients before calculating the SPOD vectors.
}

Given the forcing vector, $\hat{\mathbf{f}}$, its RESPOD mode is given by,
\begin{align} \label{eq:espodmode}
\bm{\chi}_p=\frac{E\{\hat{\mathbf{f}}\,a_p^H\}}{\uplambda_p}.
\end{align}
The RESPOD mode, $\bm{\chi}_p$, provides the forcing mode associated with the $p$th SPOD mode of the response. In \citet{boree_eif_2003}, it was shown that the extended modes can be used to isolate, from the target event, the part that is correlated with the observed coherent structure. Given, for instance, the rank-1 representation of $\hat{\mathbf{q}}$ as $\hat{\mathbf{q}}_1=a_1\bm{\psi}_1$, the corresponding RESPOD mode, $\bm{\chi}_1$, allows identification of $\hat{\mathbf{f}}_1\triangleq a_1\bm{\chi}_1$, which is the forcing correlated with $\hat{\mathbf{q}}_1$. This can be seen by comparing the two cross-covariances between the forcing and the response given as,
\begin{align}
\begin{split}
E\{\hat{\mathbf{f}}\,\hat{\mathbf{q}}_1^H\} &= E\{\hat{\mathbf{f}}a_1^H\bm{\psi}_1^H\} \\
&= E\{\hat{\mathbf{f}}a_1^H\}\bm{\psi}_1^H \\
&= \uplambda_1\bm{\chi}_1\bm{\psi}_1^H,
\end{split}
\end{align}
and,
\begin{align}
\begin{split}
E\{\hat{\mathbf{f}}_1\hat{\mathbf{q}}_1^H\} &= E\{a_1\bm{\chi}_1a_1^H\bm{\psi}_1^H\} \\
&= \bm{\chi}_1E\{a_1a_1^H\}\bm{\psi}_1^H \\
&= \uplambda_1\bm{\chi}_1\bm{\psi}_1^H.
\end{split}
\end{align}
 This indicates that $\hat{\mathbf{f}}_1$ contains all the forcing that is correlated with $\hat{\mathbf{q}}_1$. 

As the target event we consider here is the forcing term in the resolvent framework, which satisfies \eqref{eq:modifres}, substituting \eqref{eq:modifres} into \eqref{eq:psialt}, we can show that,
\begin{align} \label{eq:espodres}
\begin{split}
\bm{\psi}_p&=\frac{E\{\tilde{\mathbf{R}}\hat{\mathbf{f}}\,a_p^H\}}{\uplambda_p}, \\
&=\tilde{\mathbf{R}}\frac{E\{\hat{\mathbf{f}}\,a_p^H\}}{\uplambda_p}, \\
&=\tilde{\mathbf{R}}\bm{\chi}_p,
\end{split}
\end{align}
indicating that the correlated response and force modes are dynamically consistent, i.e., they satisfy \eqref{eq:modifres}. 

As mentioned earlier, the relation between forcing and response becomes bijective for a non-singular resolvent operator, i.e., for a given response, $\bm{\psi}$, there is a unique forcing mode that satisfies \eqref{eq:espodres} and \eqref{eq:frcinv}. This implies that for the non-singular resolvent operator, the forcing mode identified by RESPOD becomes identical to the RBE mode predicted using \eqref{eq:frcinv}. But for the case of a singular resolvent operator, RESPOD finds both the minimal-norm forcing (also predicted by RBE) necessary to generate the SPOD mode \textit{and} the correlated-but-silent components of the non-linear scale interactions.  

\subsection{Comparison of RBE and RESPOD using a simple model} \label{subsec:toy} 
As a toy model to illustrate the techniques, we consider two rank-3 resolvent operators, $\mathbf{R}_f$ and $\mathbf{R}_s$, that are, respectively, full-rank and singular,
\begin{align}
\mathbf{R}_f=\begin{bmatrix}
3 & 0 & 0 \\
0 & 2 & 0 \\
0 & 0 & 1
\end{bmatrix};\hspace{0.3cm}
\mathbf{R}_s=\begin{bmatrix}
3 & 0 & 0 & 0 & 0 \\
0 & 2 & 0 & 0 & 0 \\
0 & 0 & 1 & 0 & 0
\end{bmatrix}.
\end{align}
Both systems are driven by random forcing vectors, $\hat{\mathbf{f}}_f=[c_1,\,c_2,\,c_3]^\top$ and $\hat{\mathbf{f}}_s=[c_1,\,c_2,\,c_3,\,3c_1,\,c_4]^\top$, respectively, where $c_i$ is a random number with zero mean and unit variance. The random forcing realisations are generated using Matlab random number generator. The random number seeding is reinitialised for each simulation to ensure using the same random value series, which makes the results repeatable. The fourth and the fifth components in $\hat{\mathbf{f}}_s$ project on to the null space of $\mathbf{R}_s$, but the fourth component is fully correlated to the first since both contain the same random variable, $c_1$. Reinitialisation of the random number generator ensures that the first three components of $\hat{\mathbf{f}}_f$ and $\hat{\mathbf{f}}_s$, those that are observable through the corresponding resolvent operators, are identical. The responses obtained for the two systems are thereby identical. Two tests are carried out using 10 and 500 realisations, respectively. The leading SPOD mode for both singular and non-singular systems is obtained as $\bm{\psi}= [-0.9622, \, 0.2576, \, -0.0890]^\top$ when 10 realisations are considered and $\bm{\psi}= [-0.9943, \, -0.1056, \, 0.0148]^\top$ with 500 realisations. As the forcing in the full-rank model problem is white, i.e., $\mathbf{P}=\mathbf{I}$, by inspection, one can see that the expected value for the leading SPOD mode is $\bm{\psi}=[1,\,0,\,0]^\top$ (or its negative).  The predictions based on realisations converge to the expected value with an algebraic convergence. 

Using 10 realisations, the forcing modes computed using RBE and RESPOD, respectively $\bm{\phi}_{f/s}$ and $\bm{\chi}_{f/s}$, are, for the full-rank system,
\begin{align}
\bm{\phi}_f=\begin{bmatrix}
   -0.3207 \\
    0.1288 \\
   -0.0890
\end{bmatrix}, \textrm{ and }
\bm{\chi}_f=\begin{bmatrix}
   -0.3207 \\
    0.1288 \\
   -0.0890
\end{bmatrix},
\end{align}
while for the singular system we obtain,
\begin{align}
\bm{\phi}_s=\begin{bmatrix}
   -0.3207 \\
    0.1288 \\
   -0.0890 \\
    0 \\
    0
\end{bmatrix}, \textrm{ and }
\bm{\chi}_s=\begin{bmatrix}
   -0.3207 \\
    0.1288 \\
   -0.0890 \\
   -0.9622 \\
    0.0283
\end{bmatrix}.
\end{align}
The same results for the cases based on 500 realisations are given as,
\begin{align}
\bm{\phi}_f=\begin{bmatrix}
   -0.3314 \\
   -0.0528 \\
    0.0148
\end{bmatrix}, \;
\bm{\chi}_f=\begin{bmatrix}
   -0.3314 \\
   -0.0528 \\
    0.0148
\end{bmatrix}, \;
\bm{\phi}_s=\begin{bmatrix}
   -0.3314 \\
   -0.0528 \\
    0.0148 \\
    0 \\
    0
\end{bmatrix}, \textrm{ and }
\bm{\chi}_s=\begin{bmatrix}
   -0.3314 \\
   -0.0528 \\
    0.0148 \\
    -0.9943 \\
    0.0066
\end{bmatrix}.
\end{align}
As expected, the results obtained using RBE and RESPOD are identical in the full-rank system. The 4th component of $\bm{\chi}_s$ is three times the 1st component, where this correlation information was included in the definition of $\hat{\mathbf{f}}_s$. Although the RBE method, which finds the minimal-norm forcing for a given response, and the RESPOD method, which finds the correlated forcing for a given response, ask different questions, their results are the same when the resolvent operator is full-rank. This is because, for a full-rank resolvent operator, no silent forcing exists.
 When different numbers of realisations are considered, RBE and RESPOD yield identical results when the resolvent operator is full-rank; this is because the SPOD response mode and the RESPOD forcing mode have identical convergence in that case. This will be the case when the two methods are based on the exact same realisations, and the realisations are consistent, i.e., they satisfy \eqref{eq:modifres}. The results may differ if the database contains noise on the response and/or the forcing, which is out of the scope of this study. 
 
For the singular case, the forcing predictions obtained using RBE and RESPOD differ in the silent forcing components. RBE, by construction, predicts zero forcing in these components while RESPOD captures the correlated information between the first and the fourth component, while the fifth component, which is uncorrelated with the response, tends to zero. In a more complicated system, involving turbulent flow for example, this additional information obtained from RESPOD may be expected to provide additional insight regarding the mechanisms that underpin the forcing.

Note that, although the SPOD modes are unit vectors in the Euclidean norm, the associated forcing modes obtained using RESPOD or RBE are not unit vectors. In equation \eqref{eq:espodmode}, we see that the RESPOD mode is normalised using the response eigenvalue, and so a unit norm is not ensured. Similarly, equation \eqref{eq:frcinv} implies that, for a unit SPOD vector, the associated forcing mode obtained using RBE is not necessarily a unit vector. The norm of the forcing mode obtained by RBE/RESPOD is related by the associated gain. In RBE, amplified forcing modes will have a smaller-than-unity norm, such that $\mathbf{R}\bm{\phi}$ will be unity. In RESPOD, besides the associated gain, the norm also depends on the correlated-but-silent parts.

\section{Identifying wall-attached structures and their forcing in turbulent channel flow} \label{sec:channel}

We use the tools outlined above to analyse turbulent channel flow, our objective being to educe, from DNS data, wall-attached structures and the forcing by which they are driven. We investigate if these wall-attached structures are reminiscent of attached eddies discussed widely in the literature. 

Attached eddies begin to dominate the 1D energy spectrum of a wall-bounded flow only beyond a certain value of $Re$ ($Re_\tau\sim100,000$) \citep{perry_jfm_1986,chandran_jfm_2017}. But it has been show that structure-eduction techniques can be used to identify such structures at lower $Re$: structure functions were used by \citet{davidson_pof_2006,davidson_jfm_2006} and \citet{agostini_prf_2017}, while \citet{hellstrom_jfm_2016} and \citet{hellstrom_rsta_2017} used POD. In our study, we analyse structures that are correlated to the wall shear in a turbulent channel flow at $Re_\tau=543$, our objective being to establish: (1) if these exhibit self-similar features consistent with attached-eddies; and, (2) if the associated forcing, identified using both RBE and RESPOD, also exhibits a self-similar organisation. The underlying aim is to lay the groundwork for a resolvent-based, dynamic attached-eddy model, as discussed in the introduction. 

{In the work of \cite{yang_jfm_2017}, wall-shear in the flow direction was shown to manifest self-similarity along the log layer, reminiscent of attached eddies, and this was used to build a model for momentum cascade. Here, we use instead the wall-shear in the spanwise direction as the reference quantity to detect wall-attached structures. Our attempt to use the streamwise wall-shear did not yield clear self-similarity in the associated coherent structures, and thus is not reported here.} 

\subsection{Database and definitions}

\begin{table}
\caption{Details of the DNS database}
\label{tab:dns}
\vspace{0.3cm}
\centering
\begin{tabular}{c c c c c c c c c c}
$Re_\tau$ & $Re_{bulk}$ & $T_{max}^+$ & $\Delta t^+$ & $L_x\times L_z$ & $N_x\times N_y\times N_z$ & $\Delta x^+$ & $\Delta z^+$ & $\Delta y_{min}^+$ & $\Delta y_{max}^+$ \\
\hline
543 & 10000 & 8.84$\times10^3$ & 2.95 & $2\pi\times\pi$ & $384\times257\times384$ & 8.88 & 4.44 & 4.09$\times10^{-2}$ & 6.66
\end{tabular}
\end{table}

The flow data are provided by direct numerical simulation (DNS) of the incompressible Navier-Stokes equations using the `ChannelFlow' code of \citet{gibson_2019}. The DNS details are provided in Table \ref{tab:dns}, where $T_{max}$ and $\Delta t$ denote the total simulation time and the time step, respectively, and $L_{x/y/z}$ and $N_{x/y/z}$ denote the domain and the grid size, respectively, and the superscript $+$ denotes near-wall unit. For dealiasing, a larger number of Fourier modes (3/2 times $N_x$ and $N_z$) was used in the simulations. The subscripts $x$, $y$ and $z$ denote the streamwise, wall-normal and the spanwise directions, respectively. The mean velocities and the root-mean-square values of the DNS data used were compared against the literature \citep{alamo_pof_2003} in \citet{morra_jfm_2021}, and are therefore not presented here. {To show the statistical convergence of the simulation, we check the balance in the mean momentum equation given as, 
\begin{align} \label{eq:mombalance}
0 = 1/Re_\tau + \frac{d^2U_x^+}{d{y^{+}}^2}-\frac{\overline{uv}^+}{dy^+},
\end{align}
where $U_x$ denotes the mean streamwise velocity, $u$ and $v$ denote the velocity fluctuations in the streamwise and wall-normal directions, respectively, overbar indicates temporal averaging and the superscript $+$ denotes quantities in wall units (see \cite{wei_jfm_2005} for further details). Integrating \eqref{eq:mombalance} in $y^+$ yields
\begin{align} \label{eq:mombalanceint}
-y =  \frac{dU_x^+}{d{y^{+}}}-\overline{uv}^+.
\end{align}
The terms in \eqref{eq:mombalanceint} are plotted in figure \ref{fig:mombalance}, where it is seen that the summation of ${dU_x^+}/{d{y^{+}}}$ and $-\overline{uv}^+$ satisfies \eqref{eq:mombalanceint}, and hence the momentum balance is reached.
\begin{figure}
  \centerline{\resizebox{\textwidth}{!}{\includegraphics{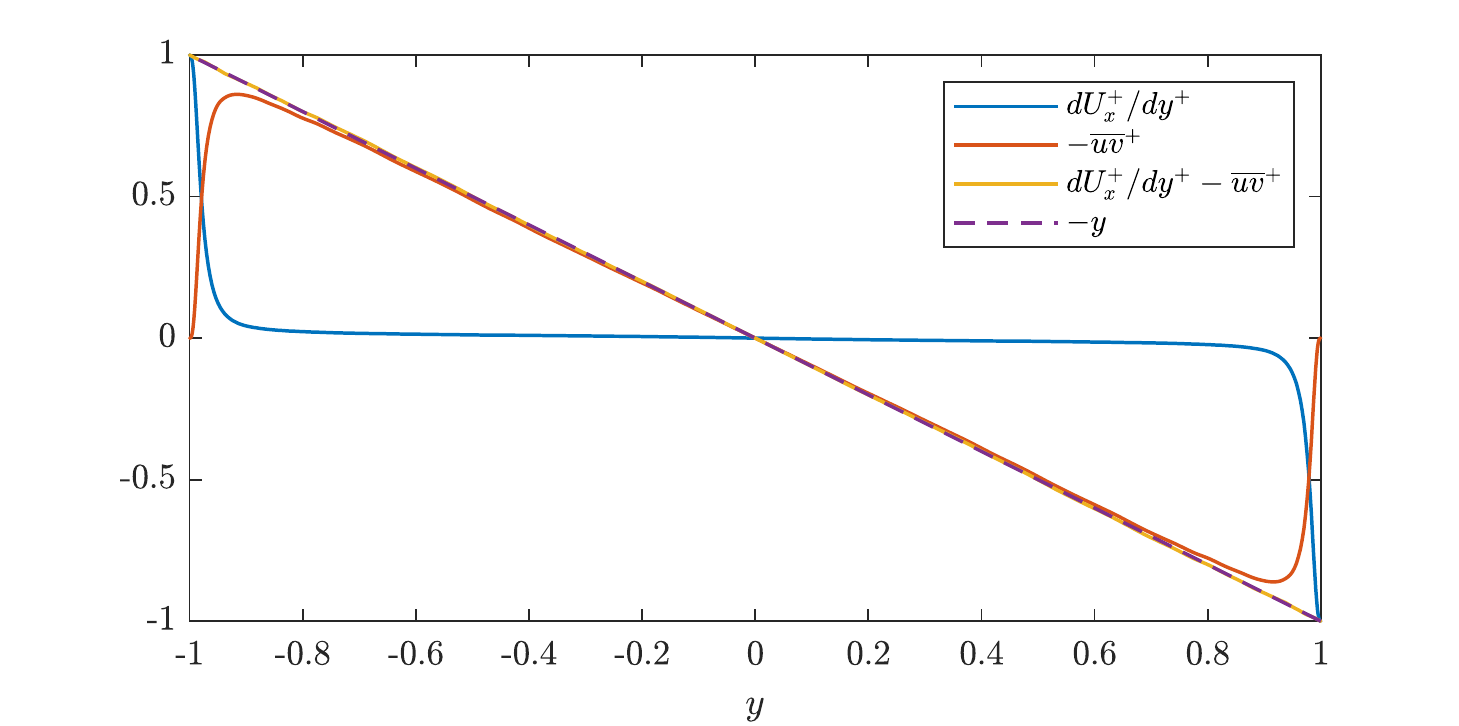}}}
  \caption{Momentum budget plot for the DNS of the channel flow.}
\label{fig:mombalance}
\end{figure} }

 The Fourier realisations are calculated using 128 fast Fourier transform (FFT) points 
with 80\% overlap between successive time blocks. The second-order exponential windowing function given in \citet{martini_arxiv_2019} is used to perform the Fourier Transform. The forcing correction associated with the windowing function is added to the Fourier realisations of the forcing as described in \citet{martini_arxiv_2019} to ensure that the forcing and the response are accurately related to each other via the resolvent operator, as shown by \cite{morra_jfm_2021} and \cite{nogueira_jfm_2021}. 

The resolvent framework for incompressible, isothermal, viscous flows is given as follows. Defining $\mathbf{u}_{tot}=\mathbf{U}+\mathbf{u}$, and $p_{tot}=P+p$, where $\mathbf{U}=[U_x,\, 0,\, 0]^T$ and $\mathbf{u}=[u,\, v,\, w]^T$ are the mean and fluctuation velocities, respectively, defined in Cartesian coordinates ordered in streamwise, wall-normal and spanwise directions, and $P$ and $p$ are the mean and fluctuating pressure, the governing equations for momentum fluctuations are given as,
\begin{align} \label{eq:ns}
\begin{split}
\partial_t\mathbf{u}+(\mathbf{U}\cdot\nabla)\mathbf{u}+(\mathbf{u}\cdot\nabla)\mathbf{U}&=-\nabla p + \frac{1}{Re}\nabla^2\mathbf{u}+\mathbf{b}+\mathbf{f},\\
\nabla\cdot\mathbf{u}&=0,
\end{split}
\end{align}
where $Re=U_{bulk}h/\nu$ denotes the Reynolds number, $\nu$  is the molecular viscosity, $\mathbf{f}=-(\mathbf{u}\cdot\nabla)\mathbf{u}$ and $\mathbf{b}=-\nabla P + Re^{-1}\nabla^2\mathbf{U}-(\mathbf{U}\cdot\nabla)\mathbf{U}$. Spatial derivative operators are given as $\nabla=[\partial_x,\, \partial_y,\, \partial_z]^T$ and $\nabla^2=\nabla\cdot\nabla$. Defining the state vector, $\mathbf{q}=[u,\,v,\,w,\,p]^T$, and taking the Fourier transform in all homogeneous dimensions, which are $x$, $z$ and $t$, with the ansatz, $\hat{\mathbf{q}}(\alpha,y,\beta,\omega)e^{i(\alpha x + \beta z - \omega t)}$, the governing equations given in \eqref{eq:ns} can be written in the matrix form as,
\begin{align} \label{eq:nsmat}
i\omega\mathbf{M}\hat{\mathbf{q}}-{\mathbf{A}}\hat{\mathbf{q}}=\mathbf{B}\hat{\mathbf{f}},
\end{align}
considering that $\omega$, $\alpha$ and $\beta$ are not simultaneously zero, such that the $\mathbf{b}$ term, which only varies in $y$, has no contribution. Note that taking the Fourier transform in $x$ and $z$ allows an accurate and inexpensive explicit construction of the resolvent operator using pseudo-spectral methods. 

Defining the transformation matrix $\hat{\mathbf{u}}=\mathbf{C}\hat{\mathbf{q}}$, we can write \eqref{eq:nsmat} in resolvent form as,
\begin{align}\label{eq:nsres}
\hat{\mathbf{u}}=\mathbf{R}\hat{\mathbf{f}},
\end{align}
where $\mathbf{R}=\mathbf{C}(-i\omega\mathbf{M}-\mathbf{A})^{-1}\mathbf{B}$ and hat denotes Fourier-transformed variables. The matrices $\mathbf{A}$, $\mathbf{B}$, $\mathbf{C}$ and $\mathbf{M}$ are given in explicit form in Appendix \ref{app:1}.

To investigate the wall-attached structures, we will define a measurement matrix,
\begin{align}
\tilde{\mathbf{C}}=[0\; 0\; \partial_z|_{y=0}\; 0],
\end{align}
which yields, when we left-multiply it with \eqref{eq:nsres},
\begin{align}
\tau_z=\tilde{\mathbf{C}}\mathbf{R}\hat{\mathbf{f}},
\end{align}
where $\tau_z=\partial_zw|_{y=0}$ is the spanwise wall shear. We choose to measure $\tau_z$ considering that it is strongly associated to quasi-streamwise vortices. We then apply the RESPOD to obtain the forcing that is correlated with the wall shear. Note that since $\tau_z$ is a scalar quantity, it has one SPOD mode that is also scalar, and the associated forcing is bound to be rank-1, which yields,
\begin{align}
\psi=\tilde{\mathbf{C}}\mathbf{R}\bm{\chi}.
\end{align}

The RESPOD mode, $\bm{\chi}$, when multiplied with the resolvent operator, $\mathbf{R}$, yields the velocity field that is correlated with $\tau_z$. The proof is provided in the following. Using the velocity vector, $\hat{\mathbf{u}}$, as the target event in \eqref{eq:espodmode} instead of the forcing, $\hat{\mathbf{f}}$, one can obtain the velocity field, $\bm{\xi}$ that is correlated with $\tau_z$ as,
\begin{align} \label{eq:uespod}
\bm{\xi}=\frac{E\{\hat{\mathbf{u}}a^H\}}{\uplambda},
\end{align}
where, for the particular case of $\tau_z$ being a scalar, $a=\langle\tau_z,1\rangle$ and the eigenvalue, $\uplambda$ is simply obtained by $E\{\tau_z\tau_z^H\}$. Substituting \eqref{eq:nsres} into \eqref{eq:uespod} reads,
\begin{align}
\bm{\xi}=\mathbf{R}\bm{\chi},
\end{align}
which proves the above statement. Since $\bm{\xi}$, also rank-1, corresponds to all and the only parts in $\hat{\mathbf{u}}$ that are correlated with $\tau_z$ (see the discussion in \S\ref{subsec:espod} or alternatively \cite{boree_eif_2003}), we consider it as the wall-attached part of $\hat{\mathbf{u}}$.

\subsection{Power spectral densities}
\begin{figure}
  \centerline{\resizebox{\textwidth}{!}{\includegraphics{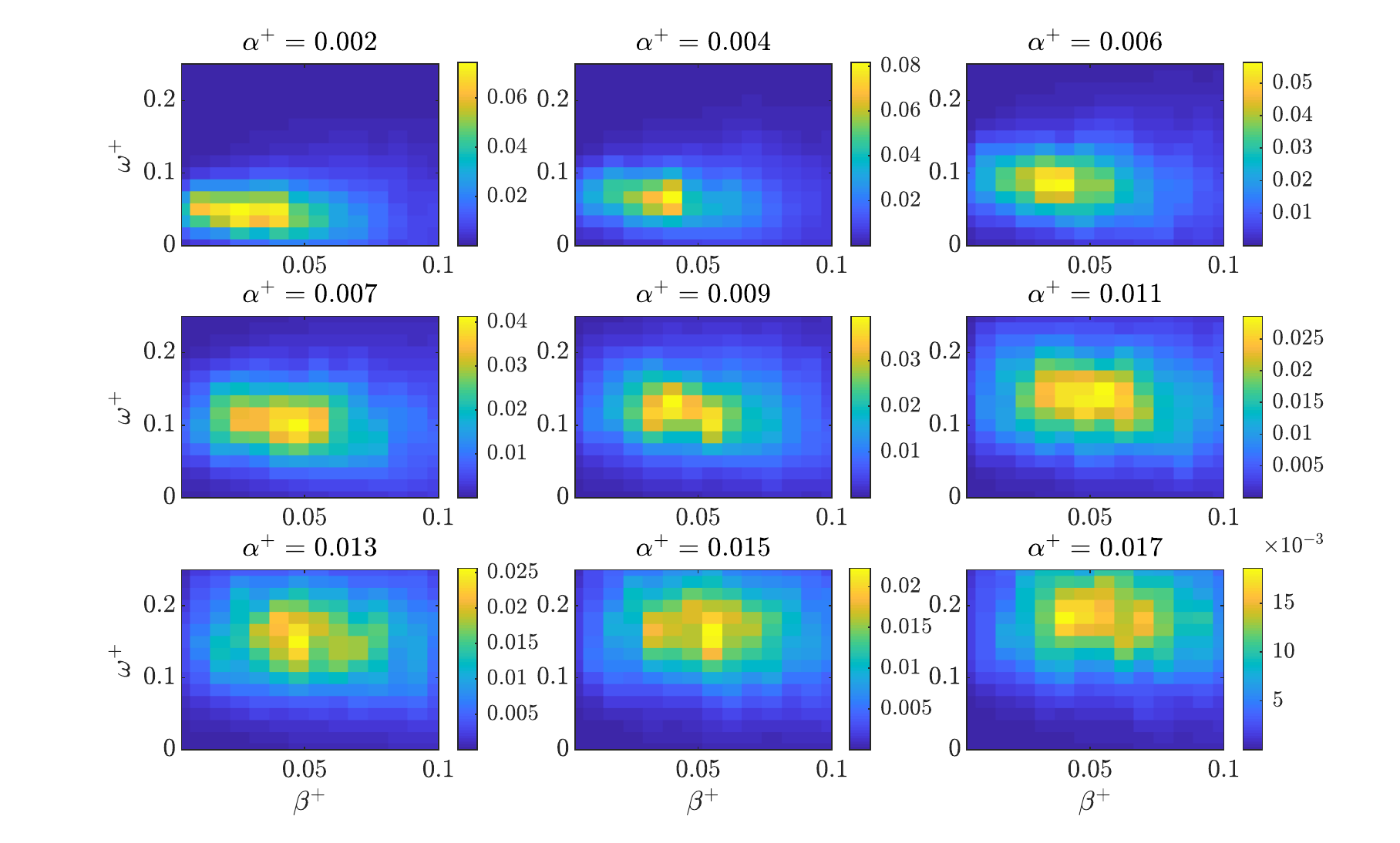}}}
  \caption{PSD map of $\tau_z$ for different $\alpha^+$ values.}
\label{fig:tauzmap}
\end{figure}

The DNS database is decomposed into Fourier modes in streamwise and spanwise directions. In figure \ref{fig:tauzmap}, we show the power spectral density (PSD) of $\tau_z$ at different $\alpha^+$ and $\beta^+$ values. Peak energy location is seen to move towards higher $\omega$ and $\beta$ for increasing $\alpha$ while the amplitude reduces. Attached eddies are expected to be the main energy containing structures in the flow \citep{townsend_jfm_1976} and exhibit a self-similar organisation. {Self-similarity implies that structures in the same hierarchy should have constant streamwise-to-spanwise and streamwise-to-wall-normal ratios. In a flow database that is decomposed into Fourier modes in the streamwise and the spanwise directions, self-similarity can be investigated by imposing it in the horizontal plane by fixing the streamwise-to-spanwise aspect ratio (AR) and searching for it in the wall-normal direction, similar to \citet{hellstrom_jfm_2016} and \cite{hellstrom_rsta_2017}. In what follows, we first set $\lambda_x^+/\lambda_z^+=6$, where $\lambda_x$ and $\lambda_z$ are the streamwise and spanwise wavelengths, and look for self-similar structures at this AR. We then extend our analysis to a range of ARs and provide an overall view of self-similarity for the given channel flow. We first investigate self-similarity of structures at frequencies with high spectral energy which is followed by investigation of self-similarity in less energetic structures.}

\begin{figure}
  \centerline{\resizebox{\textwidth}{!}{\includegraphics{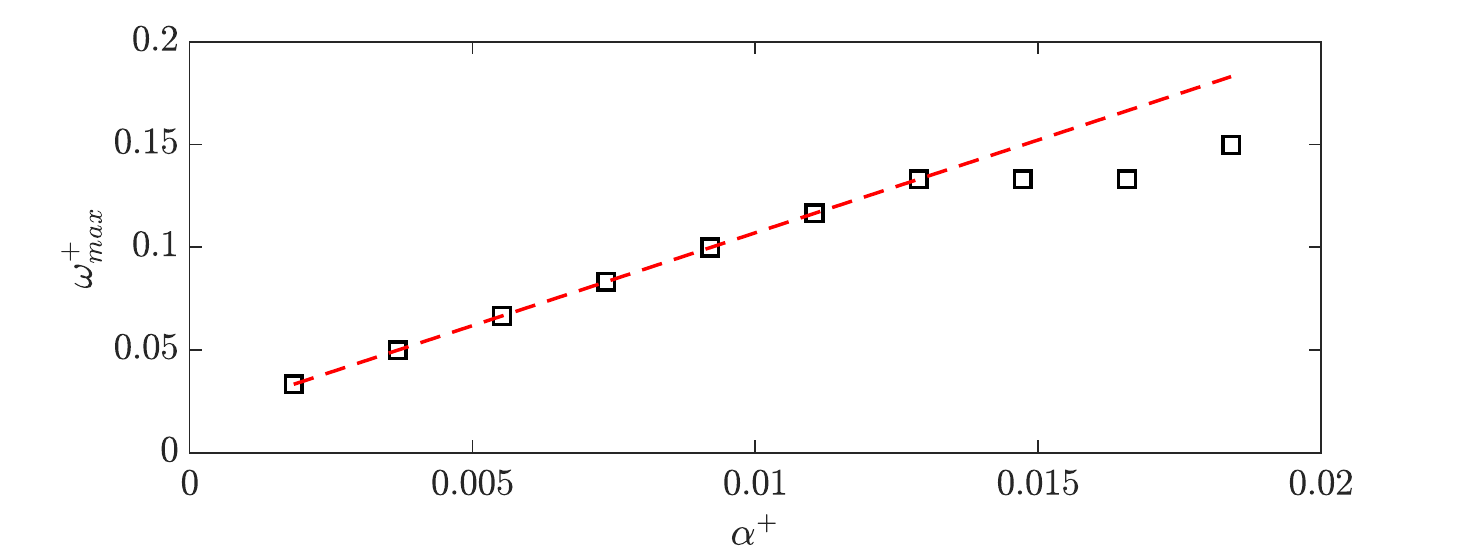}}}
  \caption{Maximum-energy-containing frequency, $\omega_{max}$ of spanwise wall shear, $\tau_z$ for different Fourier-mode pairs with AR=6. The red-dashed line indicates the best-fit minimizing $L_1$-norm error.}
\label{fig:omgmaxar6}
\end{figure} 

\begin{figure}
  \centerline{\resizebox{\textwidth}{!}{\includegraphics{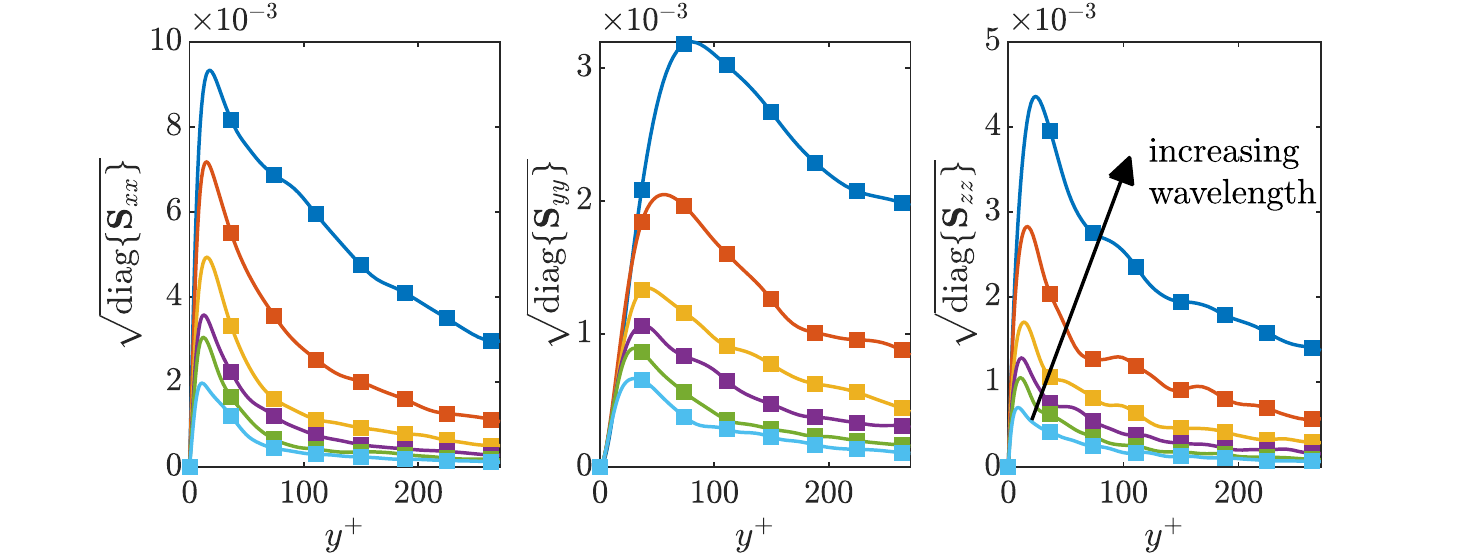}}}
  \caption{PSD of the response, $\hat{\mathbf{q}}$, (solid lines) in comparison to its resolvent-based prediction, $\mathbf{R}\hat{\mathbf{f}}$, (markers) at $\omega_{max}$ for different Fourier-mode pairs with AR=6. Wavelengths plotted are $\lambda_x^+=487$, 569, 683, 853, 1137, and 1706.}
\label{fig:dirvsresp}
\end{figure} 

\begin{figure}
  \centerline{\resizebox{\textwidth}{!}{\includegraphics{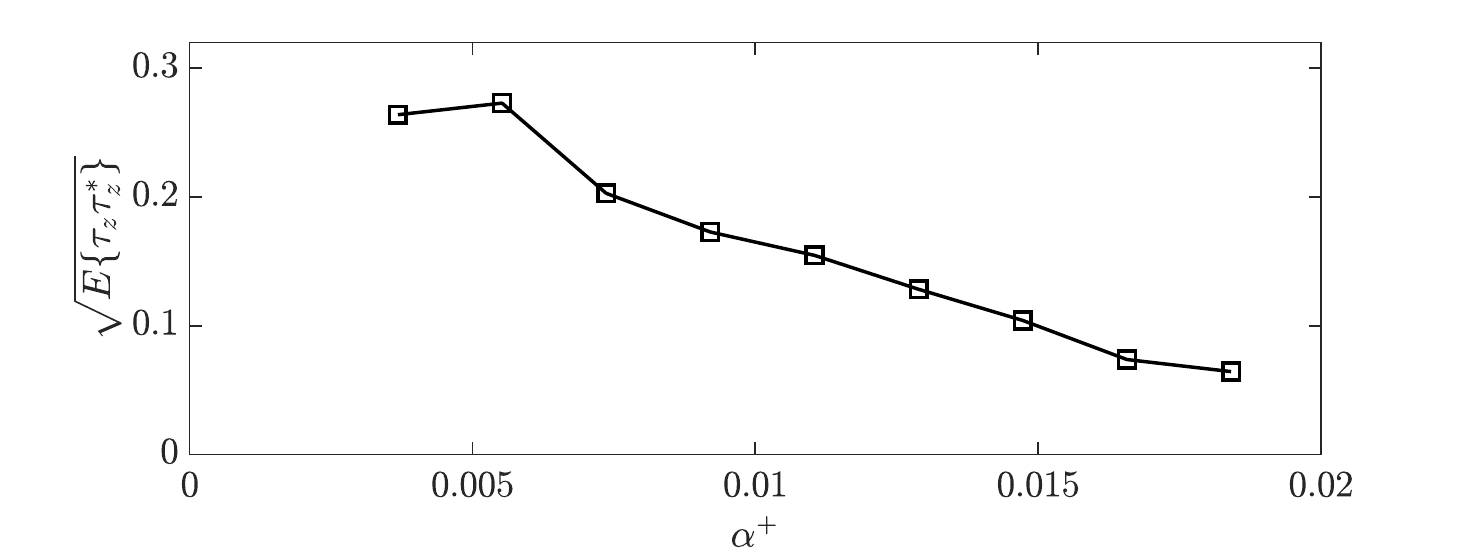}}}
  \caption{Amplitude spectral density of spanwise wall shear, $\tau_z$ at $\omega_{max}$ for different Fourier-mode pairs with AR=6.}
\label{fig:tauzasd}
\end{figure} 

Maximum-energy containing frequency, $\omega_{max}$ for different $\alpha^+$ values with AR${}=6$ is shown in figure \ref{fig:omgmaxar6}. We see that $\omega_{max}$ increases almost linearly with $\alpha^+$. We approximate this trend with the red line shown in the figure which is the best linear fit minimizing $L_1$ norm error. $L_1$ norm is chosen for the trendline to be less sensitive to outliers. The frequency corresponding to each $(\lambda_x^+,\lambda_z^+)$ pair is selected using this trend line. 

The accuracy of the forcing database in $Re_\tau=543$ flow is verified by comparing the state, $\hat{\mathbf{q}}$, to its resolvent-based prediction, $\mathbf{R}\hat{\mathbf{f}}$, in figure \ref{fig:dirvsresp}. Such an evaluation has already been performed by \cite{morra_jfm_2021} for two dominant structures, and it is here extended to a broader range of wavenumbers. As seen in the figure, the resolvent-based prediction of the response matches the DNS state data for the entire wavenumber span considered. 

The square root of PSD of the wall shear $\tau_z$ calculated at $\omega_{max}$ is shown in figure \ref{fig:tauzasd}. The amplitude is seen to decrease for increasing $\alpha^+$ as expected from figure \ref{fig:tauzmap}. The trend is almost linear for $\alpha^+>7.4\textrm{e-}3$.

\subsection{Wall-attached coherent structures and associated forcing for $AR=6$}

\begin{figure}
  \centerline{\resizebox{\textwidth}{!}{\includegraphics{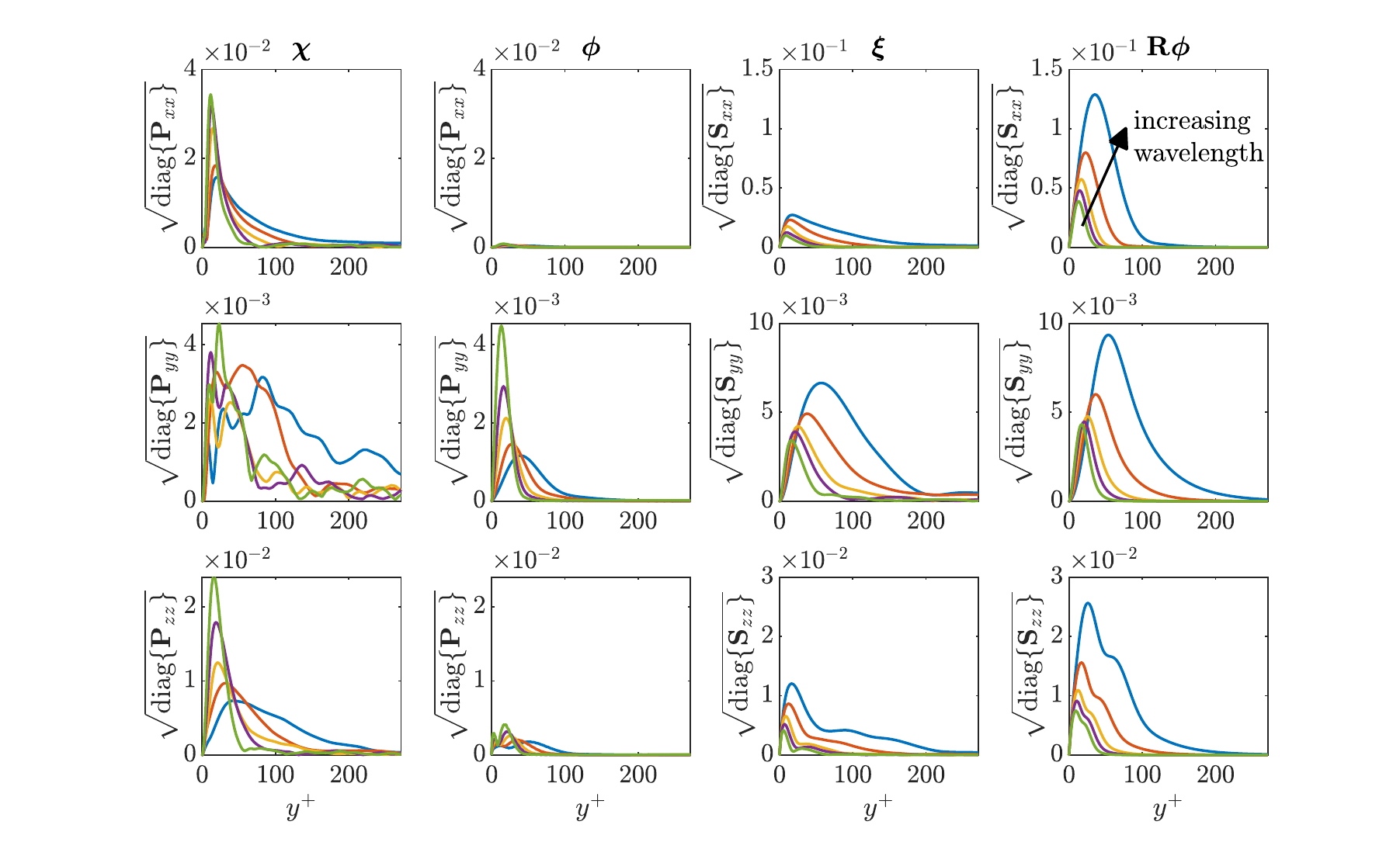}}}
  \caption{Comparison of the wall-shear-associated forcing modes obtained by RESPOD (1st column) and RBE (2nd column), respectively, and the corresponding velocity fields (3rd and 4th columns, respectively) respectively, at $\lambda_x^+=341$, 487, 682, 1137 and 1706 with $\lambda_x^+/\lambda_z^+=6$. }
\label{fig:respodvsrbe}
\end{figure}

We now focus on the wall-attached structures at the peak-energy frequency, $\omega_{max}^+$, for each $(\lambda_x^+,\lambda_z^+)$ pair with AR fixed to be 6. The modes are forced to be symmetric in $u$ and $w$ (and anti-symmetric in $v$) about the channel center in the wall-normal direction by averaging the flow fields in the bottom and the upper halves of the channel (considering a minus sign in $v$), as in \cite{abreu_ijhff_2020}. In figure \ref{fig:respodvsrbe}, wall-shear-associated forcing modes obtained using RESPOD and RBE methods, respectively, and the velocity fields generated by these forcing modes are shown. Note that the mode amplitudes are adjusted to have unit wall shear. The forcing from the RBE method, $\bm{\phi}$, shows that the spanwise and wall-normal components of forcing drive the wall-shear dynamics observed in the flow at this AR, while the streamwise component is almost not required. For the forcing from RESPOD method, $\bm{\chi}$, on the other hand, the forcing mechanisms that generate wall shear involves to a large extent the streamwise component of forcing, although its overall contribution to wall-shear dynamics can be small as suggested by RBE results. This indicates strong cancellations among the responses to different forcing components, similar to the results in \cite{morra_jfm_2021}. This will be further investigated at the end of this subsection. As discussed earlier, the response, $\bm{\xi}=\mathbf{R}\bm{\chi}$, yields the velocity field associated with the wall shear. Similarly, $\mathbf{R}\bm{\phi}$ can be interpreted as best prediction of the correlated velocity field when no data for forcing is available. This amounts to flow estimation using low-rank measurements with the assumption of forcing CSD, $\mathbf{P}=\mathbf{I}$, as discussed in \cite{martini_jfm_2020}. We see that the velocity modes associated with the wall shear are significantly overpredicted when RBE is used.

\begin{figure}
  \centerline{\resizebox{\textwidth}{!}{\includegraphics{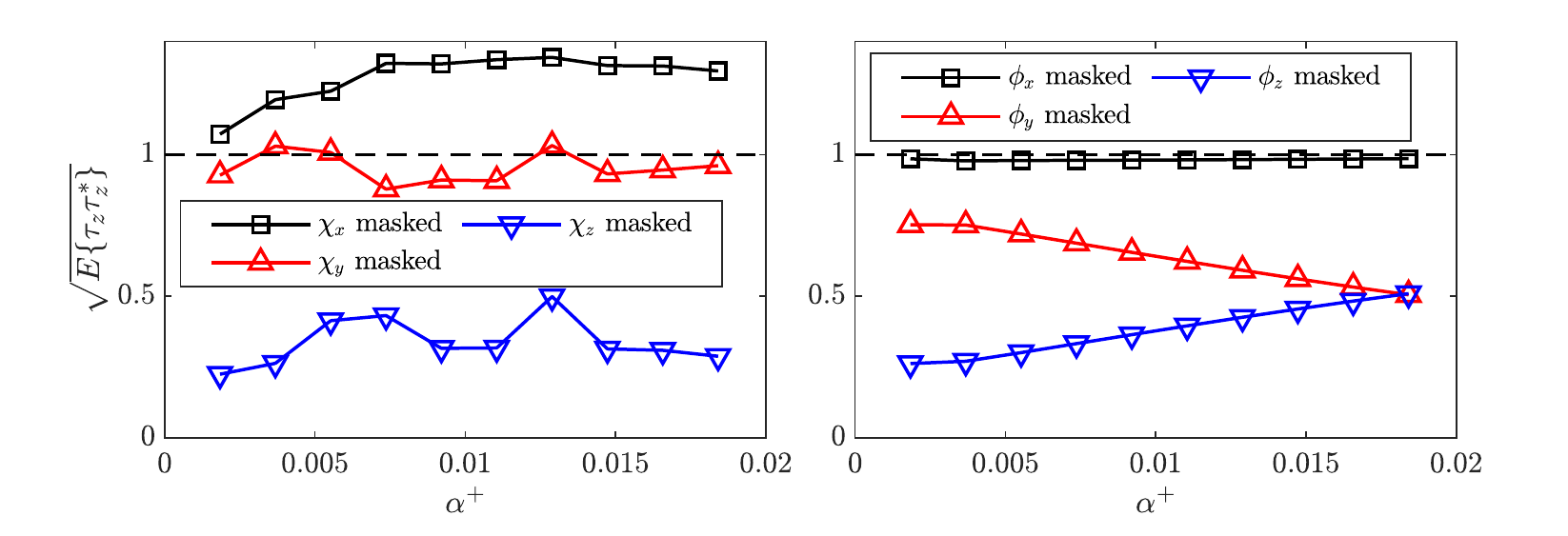}}}
  \caption{Square root of PSD of $\tau_z$ obtained using partially masked $\bm{\chi}$ (left) and $\bm{\phi}$ (right) at different wavenumbers with AR=6. Unmasked PSDs are all unity, shown by the dashed line. }
\label{fig:tauzmasked}
\end{figure}

As mentioned above, the difference between the minimal-norm forcing, $\bm{\phi}$ and the wall-shear-correlated forcing, $\bm{\chi}$, implies that contributions to $\tau_z$ from each component of $\mathbf{P}_{\chi}=\bm{\chi}\bm{\chi}^H$ cancel each other to a significant extent. To understand the effect of a particular component of $\bm{\chi}$ (or $\bm{\phi}$), we calculate the wall shear using $\bm{\chi}$ (or $\bm{\phi}$) with that component masked. The resulting wall-shear PSDs for partially masked $\bm{\chi}$ and $\bm{\phi}$, respectively, are shown in figure \ref{fig:tauzmasked}. The forcing modes are scaled such that using the full forcing mode generates a wall shear with unit amplitude at each wavenumber pair. For the wall-shear-correlated forcing, $\bm{\chi}$, we see that the wall-normal  component has the least effect on the resulting $\tau_z$ amplitude at every wavenumber pair. In the case of minimal-norm forcing, $\bm{\phi}$, on the other hand, the streamwise component has negligible effect on wall-shear dynamics. This implies that the mechanisms involved in the two forcing modes are indeed different, although amounting to the same overall response. For $\bm{\phi}$, wall shear is generated by a pure streamwise vortical forcing at every wavenumber. For $\bm{\chi}$, it is the streamwise and spanwise components which are important with the latter being the main driving component. This implies that turbulence does not follow the optimal path to generate wall-shear fluctuations. The fact that masking $\chi_x$ increases $\sqrt{E\{\tau_z\tau_z^*\}}$ implies that simultaneous presence of the streamwise and spanwise components cause some cancellation in the wall-shear fluctuation amplitude.

{Note that the predictions from the RBE method can be improved using an eddy-viscosity model. The eddy viscosity is often used to enhance modeling properties of the resolvent operator \citep{pickering_aiaa_2019,morra_jfm_2019}. It was used in \cite{towne_jfm_2020} and was more effective to recover flow ($\mathbf{q}$) statistics. As shown in \cite{morra_jfm_2021}, the eddy viscosity embeds some of the forcing statistics in the resolvent operator, but not all \citep{amaral_jfm_2021}. A drawback of using eddy viscosity is that the forcing of an eddy-viscosity-based resolvent operator do not correspond to the actual nonlinear Navier-Stokes terms. Since we are interested in these nonlinear terms, we use the molecular-viscosity-based resolvent operator as it allows to directly probe the nonlinear terms via the forcing, $\mathbf{f}$.}

\subsection{Self-similarity of wall-attached forcing and response structures for $AR=6$}

\begin{figure}
  \centerline{\resizebox{\textwidth}{!}{\includegraphics{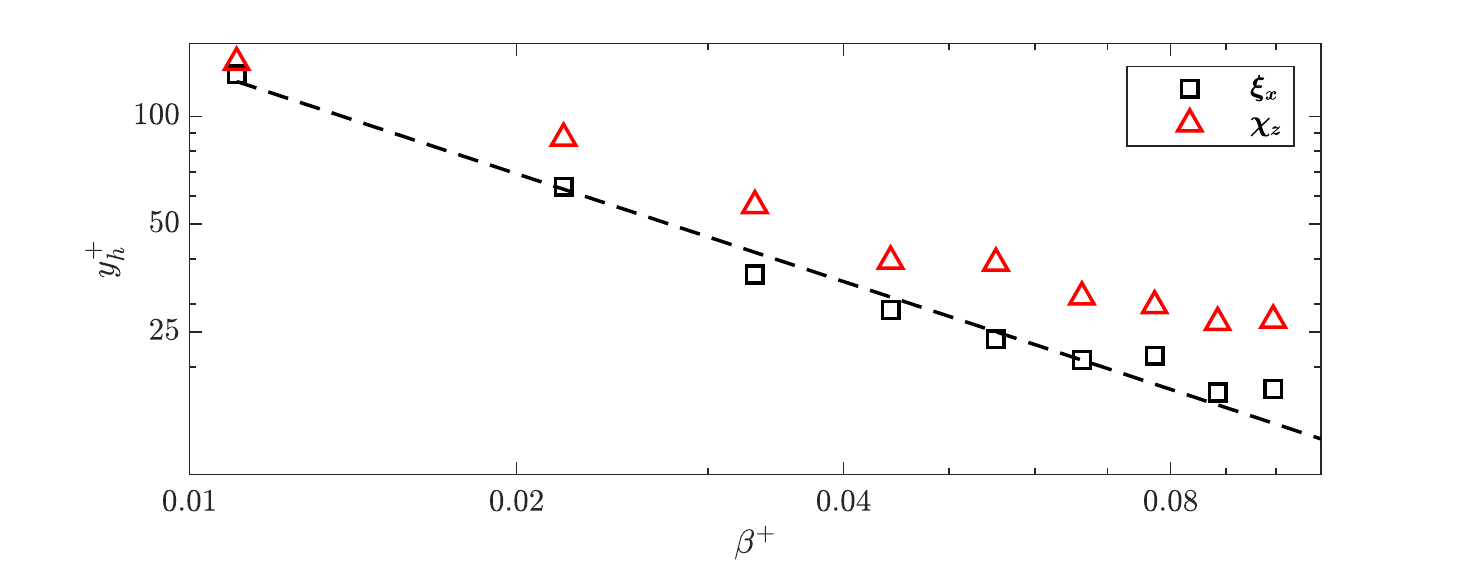}}}
  \caption{Mode center, $y_h^+$ obtained using $\xi_x$ (square) and $\chi_z$ (triangle) at different wavenumbers with AR=6. The dashed line shows $1/{\beta^+}$ trend.}
\label{fig:yhvsbeta}
\end{figure}

Townsend's AEH states that the attached eddies scale in size with respect to their distance to the wall \citep{townsend_jfm_1976,perry_jfm_1982}. We will assess the validity of this assumption for the wall-attached structures investigated in this study. As mentioned earlier, by keeping the AR fixed, we impose self-similarity in the streamwise and spanwise directions. Regarding the modes seen in figure \ref{fig:respodvsrbe}, although the amplitudes show different trends particularly for different forcing components, we see that the mode shapes exhibit to a certain extent a self-similar trend except the wall-normal component of the forcing mode $\bm{\chi}$. {The self-similarity can be made apparent by a proper scaling of the modes in the wall-normal direction and normalisation of each mode with its peak amplitude.} To assess the self similarity in the wall-normal direction, \citet{hellstrom_jfm_2016} proposed a scaling based on the peak of the POD modes. In a similar spirit, we propose a scaling based on the integral function,
\begin{align} \label{eq:modectr}
g(\alpha,y) = \int_0^{y} |{\xi}_x(\alpha,y,\beta(\alpha),\omega_{max}(\alpha))|dy,
\end{align}
where ${\xi}_x$ is the streamwise component of the wall-shear-correlated velocity mode, $\bm{\xi}$, and we use this function to define a characteristic length scale, $y_h$, such that $g(\alpha,y_h)=g(\alpha,H)/2$, where $H$ is the channel half-height. In other words, $y_h(\alpha)$ is the wall-normal position that divides the area under $|\xi_x|$ into two equal parts. The $\beta^+$ dependence of $y_h^+$ is shown in figure \ref{fig:yhvsbeta}. We repeat this analysis with ${\chi}_z$ instead of ${\xi}_x$ and show the results in the same figure. The component $\chi_z$ is chosen as it is the dominant component in generating wall shear as shown in figure \ref{fig:tauzmasked}. Consistent with the attached-eddy hypothesis \citep{townsend_jfm_1976,perry_jfm_1982}, the wall-attached structures with the same AR computed at the peak-energy frequency, $\omega_{max}$, follow a $\beta^{-1}$ trend except the very small scales ($\beta^+>0.09$). The forcing modes also follow a similar trend which again deviates from $\beta^{-1}$ at small scales. The wall-attached structures scaling with $\beta^{-1}$, consistent with the AEH, suggests a self-similar behaviour. The forcing modes following a similar trend suggests that the mechanisms that generate these structures may also be self-similar, i.e. dynamic self-similarity in addition to Townsend's kinematic self-similarity.

\begin{figure}
  \centerline{\resizebox{\textwidth}{!}{\includegraphics{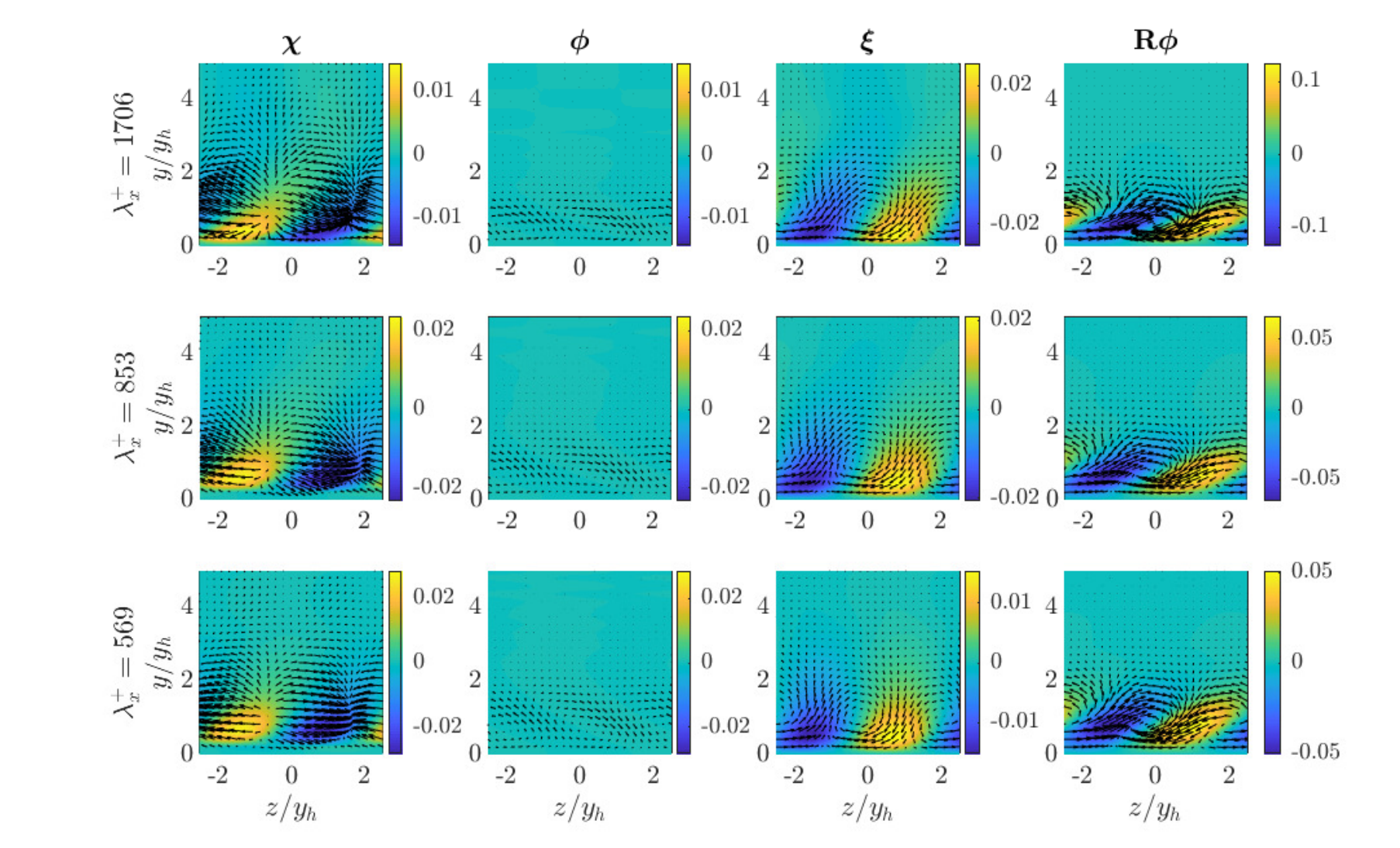}}}
  \caption{Reconstruction of wall-shear-associated forcing modes, $\bm{\chi}$ (1st column) and $\bm{\phi}$ (2nd column) and the corresponding velocity fields (3rd and 4th columns, respectively) in the $y$-$z$ plane at three wavenumbers (from top to bottom). Color-plots indicate the $x$-component while vectors show $y$- and $z$-components. The domain is rescaled using $y_h$ for each wavenumber.}
\label{fig:reconsmode}
\end{figure} 

The characteristic eddy length scale, $y_h$, can now be used to normalise the structures seen in figure \ref{fig:respodvsrbe}. We reconstruct the forcing and response modes in $y-z$ plane, which corresponds to channel cross-section. Mode phases at each $(\lambda_x^+,\lambda_z^+)$ pair are shifted to have $\angle\tau_z=0$. We use $y_h$ based on ${\xi}_x$ to scale both the forcing and response modes in $y$- and $z$-directions. We normalize the wall-normal and spanwise components of both forcing modes, $\bm{\chi}$ and $\bm{\phi}$, with ${\chi}_x$, and similarly those of the response modes with ${\xi}_x$. The reconstructed modes reveal self-similarity of the wall-attached structures and the associated forcing. The streaks and the streamwise vortices seen in the response modes, $\bm{\xi}$, indicate that the wall shear, $\tau_z$ is associated with the lift-up mechanism \citep{brandt_ejm_2014}. As discussed earlier, the RBE method overpredicts the wall-shear-correlated velocity field. The modes are once again reminiscent of lift-up mechanism with the streaks and the streamwise vortices are considerably more tilted in spanwise direction. The optimal forcing, $\bm{\phi}$, which is sufficient to create the wall shear observed in the flow, has a streamwise-vortex-like structure. This is consistent with the optimal forcing modes in Couette flow discussed in \cite{hwang_jfm_2010}. The forcing modes that actually take place in the flow on the other hand do not immediately reveal such a simple structure, although showing self-similarity in itself. 

\subsection{Extension of self-similarity analysis to other ARs}
\begin{figure}
  \centerline{\resizebox{\textwidth}{!}{\includegraphics{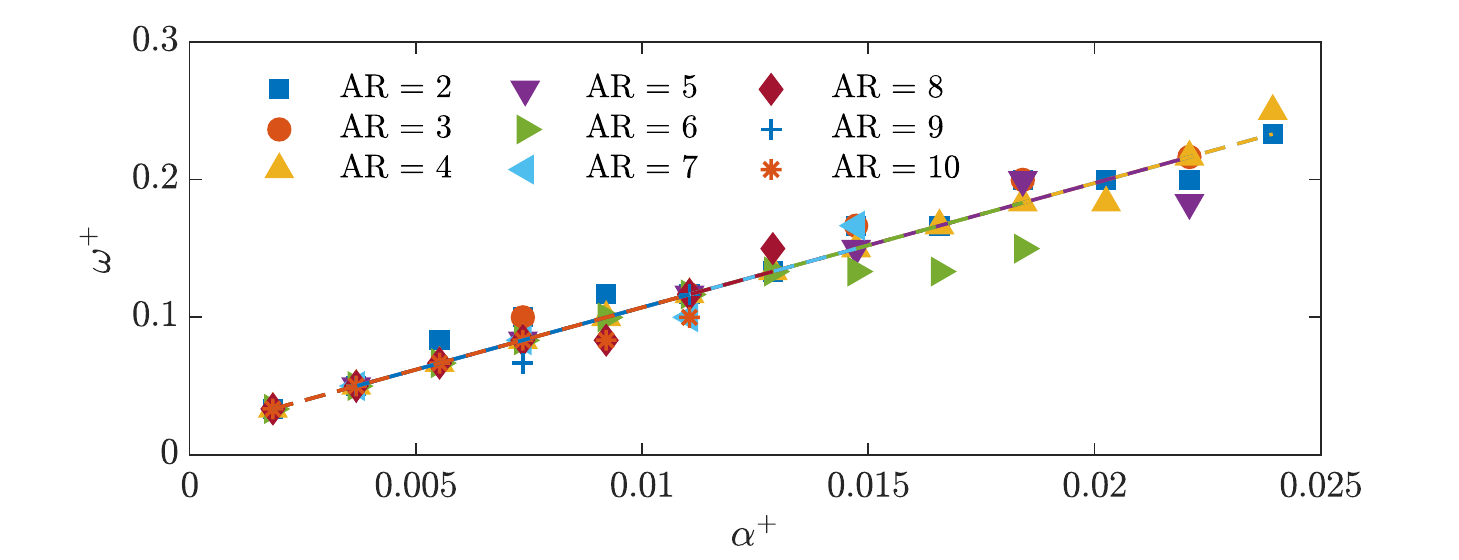}}}
  \caption{Peak-energy containing frequencies for different wave number pairs with fixed AR. Different markers indicate different AR values. The coloured dashed lines correspond to the best-fit lines for different ARs using the same color code.}
\label{fig:omgmaxall}
\end{figure}

Given the self-similarity observed at the wall-attached structures with AR equal to 6, we now extend our analysis to other ARs ranging from 2 to 10. Once again we look for the dominant structures extracted by choosing the maximum-energy-containing frequency for each $(\lambda_x^+,\lambda_z^+)$ pair for a given AR. Figure \ref{fig:omgmaxall} shows the peak-energy frequency, $\omega_{max}^+$ vs. wavenumber plots for different ARs, together with the linear trends minimizing $L_1$-norm error. We see that all the trends for different ARs collapse onto the same line given by, $\omega^+=9.04\alpha^+ + 0.0166$. We use this linear trend to set the frequency for a given $(\lambda_x^+,\lambda_z^+)$ pair at a given AR.

\begin{figure}
  \centerline{\resizebox{\textwidth}{!}{\includegraphics{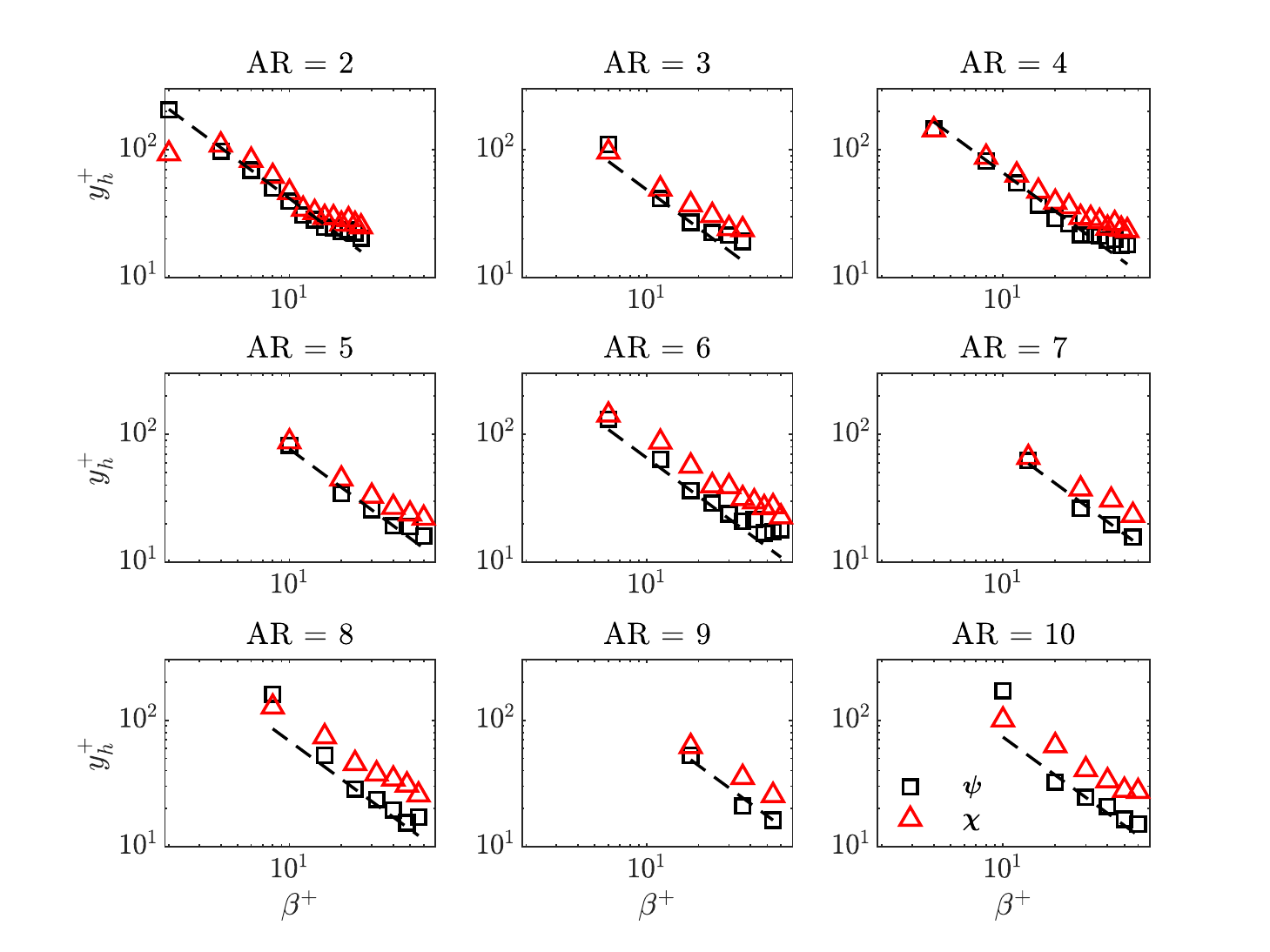}}}
  \caption{Mode center, $y_h^+$ obtained using $\xi_x$ (square) and and $\chi_z$ (triangle) at different wavenumbers. Each subplot shows results for a different AR. The dashed lines show $1/{\beta^+}$ trend.}
\label{fig:yhvsbetaall}
\end{figure} 

We calculate the mode center, $y_h$ and investigate its change with respect to the spanwise wavenumber, $\beta$. Similar to the analysis conducted for AR=6 case, we compute $y_h$ using ${\xi}_x$ and ${\chi}_z$, respectively, and show the results for different ARs in figure \ref{fig:yhvsbetaall}. Overall, we observe the $\beta^{-1}$ trend for both the response and forcing structures at every AR, which imply that self-similarity may be present for all the dominant wall-attached structures and the associated forcing for a broad range of ARs. In general, the small-scale structures near the wall are seen to deviate from the $\beta^{-1}$ trend for all the ARs. This deviation may be due to increasing viscous effects  at this region. We see that at certain ARs, the largest forcing and/or response structures also deviate from the from the $\beta^{-1}$ trend. This may be caused by these structures being affected by the other channel half at the given $Re$.

To quantify self-similarity we define the following measure. Given $\zeta$ as any component of any response or forcing mode for a given $(\lambda_x^+,\lambda_z^+)$ pair at a given AR, we first normalize $\zeta$ with its peak value given as, $\zeta_n$ and then perform $y_h$ scaling as,
\begin{align}
\zeta_n(y)=\tilde{\zeta}_n(y/y_h),
\end{align}
to get $\zeta_n$ in self-similar coordinates, where $y_h$ is calculated using ${\xi}_x$. After performing this analysis for all the wavenumber pairs at a given AR, we find the `mean' self-similar mode by averaging these modes as,
\begin{align}\label{eq:modeave}
E\{\tilde{\zeta}_n\}=\frac{1}{N}\sum_i^N\tilde{\zeta}_n^{(i)},
\end{align}
where $N$ denotes the total number of wavenumber pairs at that AR. We then compute the alignment of $\tilde{\zeta}_n^{(i)}$, which corresponds to the $i^{\textrm{th}}$ wavenumber pair, with this mean self-similar mode as,
\begin{align} \label{eq:align}
\gamma_\zeta^{(i)}=\frac{\left|\langle\tilde{\zeta}_n^{(i)},E\{\tilde{\zeta}_n\}\rangle\right|}{\sqrt{\langle\tilde{\zeta}_n^{(i)},\tilde{\zeta}_n^{(i)}\rangle}\sqrt{\langle E\{\tilde{\zeta}_n\},E\{\tilde{\zeta}_n\}\rangle}}
\end{align}
According to \eqref{eq:align}, $\gamma_\zeta^{(i)}$ becomes one in case of perfect self-similarity and zero in case of no self-similarity. {Assuming that a self-similar behaviour is present in all the data considered, the averaging process will clearly reveal the self-similar profile, with each data point showing small deviations from it. Note however that if self-similar behavior  is only found in limited a range of parameters, e.g. $\lambda_z$, including data points outside this range can mask the self-similar behaviour. That is, the approach used is prone to false negatives, being a conservative identification method.}

\begin{figure}
  \centerline{\resizebox{\textwidth}{!}{\includegraphics{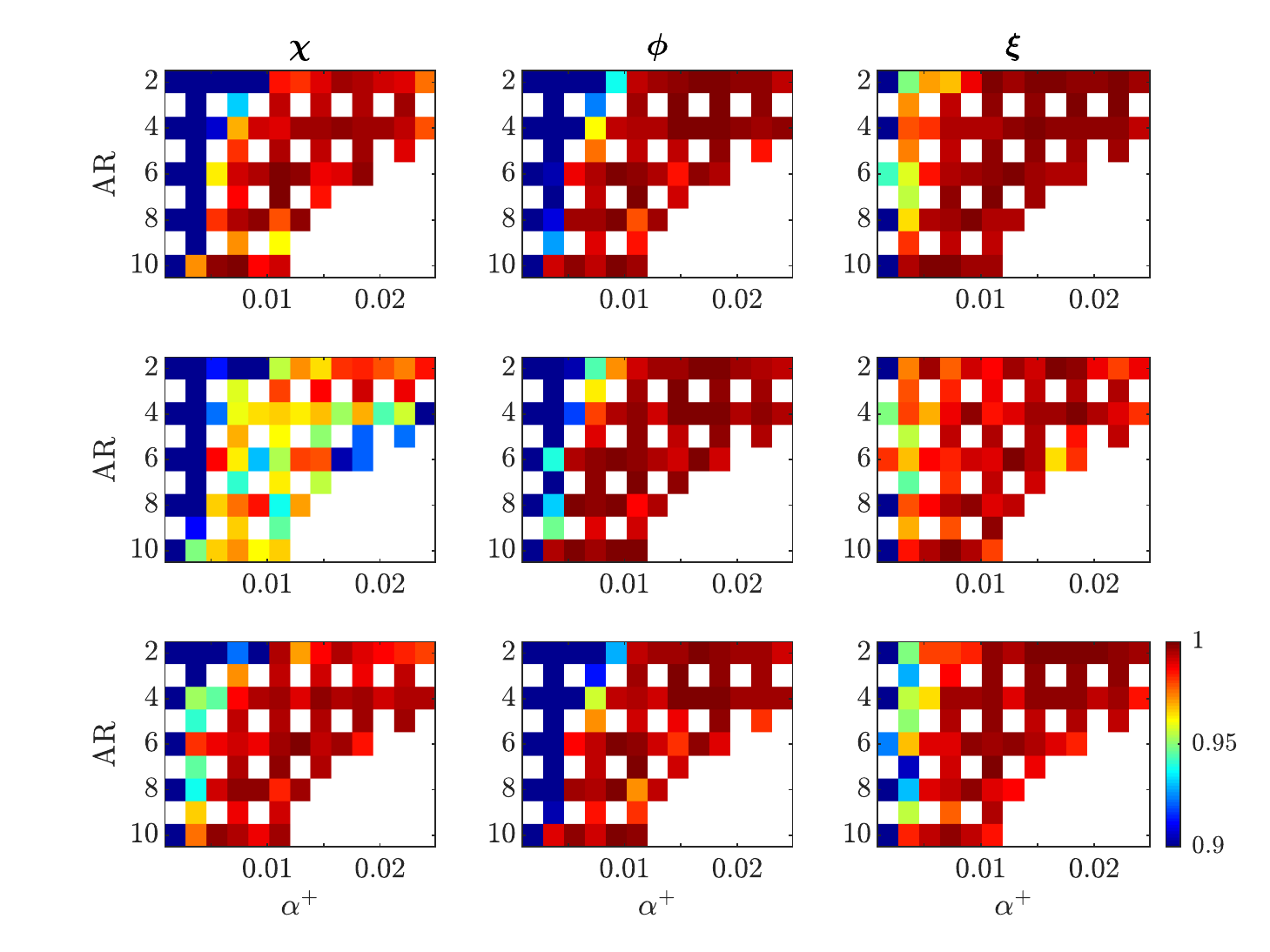}}}
  \caption{Self-similarity maps, $\gamma_{\bm{\chi}}$ (left), $\gamma_{\bm{\phi}}$ (center), and $\gamma_{\bm{\xi}}$ (right) for different ARs and $\alpha^+$ values. Rows correspond to the $x$-, $y$- and $z$-components from top to bottom.}
\label{fig:selfsimmap}
\end{figure}

In figure \ref{fig:selfsimmap}, we show the similarity maps, $\gamma_{\bm{\chi}}$, $\gamma_{\bm{\phi}}$, and $\gamma_{\bm{\xi}}$ for different ARs and $\alpha^+$ values. White regions in the figure correspond to wavenumber pairs that are not contained in the database. In general, as all the modes have similar phase and the mode shapes are never too different from each other, the self-similarity coefficient given in \eqref{eq:align} takes for almost all the cases values close to 1. However, we observed by visual inspection that self-similarity is sufficiently clear only for $\gamma>0.9$. Therefore, we set this as the lower limit in the maps shown in figure \ref{fig:selfsimmap}. The wall-attached structures given by $\bm{\xi}$ show strong self-similarity except the very large scale structures, which, once again, potentially are affected by other channel half. We see slight reduction in the self similarity of the smallest scales particularly in wall-normal and spanwise velocity components, $\xi_y$ and $\xi_z$, respectively. This reduced self-similarity is in agreement with the deviation of the small-scale structures from $\beta^{-1}$ trend as it was shown in figure \ref{fig:yhvsbetaall}. 

Parallel to the wall-attached structures, the associated minimal-norm forcing, $\bm{\phi}$ shows also self-similarity, although slightly reduced, in all its components except the large-scale structures. Note that the minimal-norm forcing, $\bm{\phi}$, involves only the response and the resolvent operator. Given that the response is self-similar, this result implies that the resolvent operator induce self similarity as well. This is in line with the findings of \cite{hwang_jfm_2010}, \cite{mckeon_jfm_2017} and {\cite{sharma_ptrsa_2017}}. 

Regarding the wall-shear-correlated forcing, $\bm{\chi}$, we observe self similarity in the streamwise and spanwise components, $\chi_x$ and $\chi_z$, respectively, with that in $\chi_z$ being slightly higher. No such self similarity is seen the wall-normal component, $\chi_y$. A potential reason to that may be the lower amplitude of forcing modes in the $y$-direction compared to other components, as illustrated in figure \ref{fig:respodvsrbe}; statistical convergece for low-amplitude components is sufficiently harder, as $\chi_y$ is the smallest and the least effective forcing component in terms of wall-shear dynamics (see figure \ref{fig:tauzmasked}). Even being limited to $\chi_x$ and $\chi_z$, given that wall shear is mainly determined by these components, the self-similarity seen in $\bm{\chi}$ implies self-similar mechanisms associated with the wall-attached structures. Presence of such self-similar mechanisms may be beneficial for modelling of wall-shear dynamics and attached eddies.

\begin{figure}
  \centerline{\resizebox{\textwidth}{!}{\includegraphics{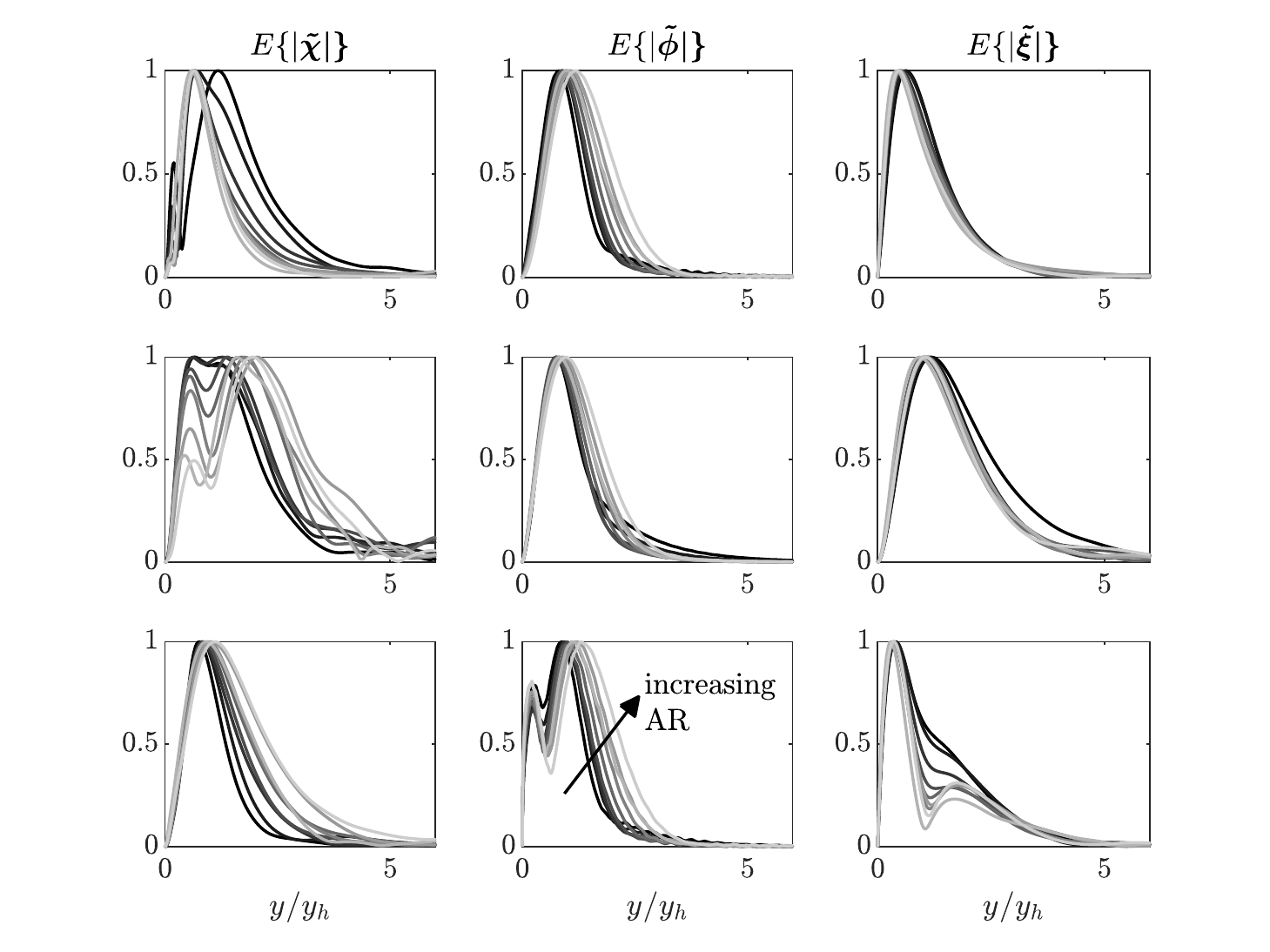}}}
  \caption{Mean self-similar modes for $\bm{\chi}$ (left), $\bm{\phi}$ (center) and $\bm{\xi}$ (right). Rows correspond to the $x$-, $y$- and $z$-components from top to bottom.}
\label{fig:expselfsim}
\end{figure} 

\begin{figure}
  \centerline{\resizebox{\textwidth}{!}{\includegraphics{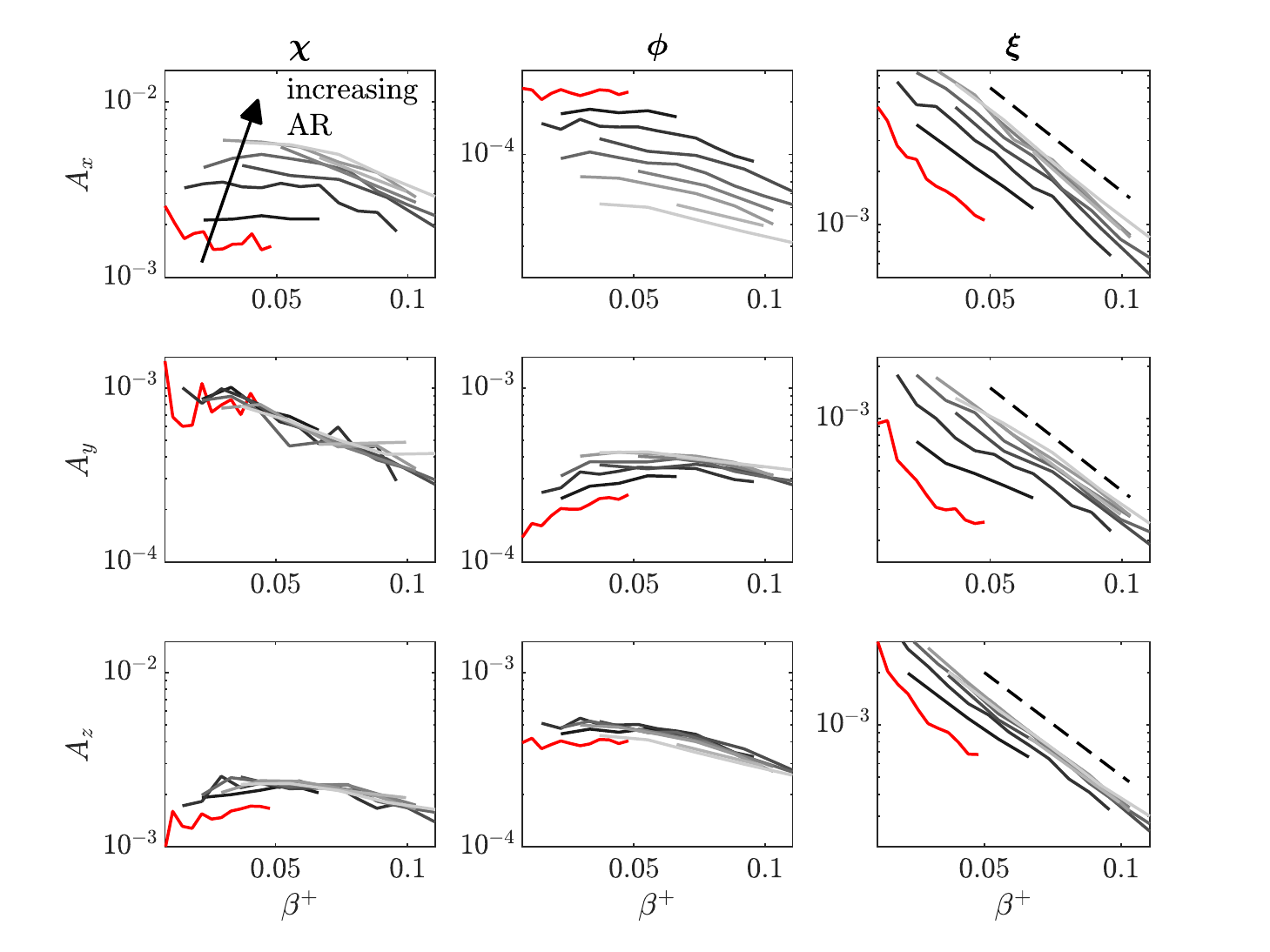}}}
  \caption{Mode amplitudes for $\bm{\chi}$ (left), $\bm{\phi}$ (center) and $\bm{\xi}$ (right) at different wavenumber pairs. Rows correspond to the $x$-, $y$- and $z$-components from top to bottom. ARs range from 2 to 10. AR=2 is shown with red and the rest with gray scale. Dashed lines indicate $1/(\beta^+)^2$ trend.}
\label{fig:modeAmp}
\end{figure} 

The mean self-similar modes of forcing and response for different ARs are shown in figure \ref{fig:expselfsim}. The streamwise and wall-normal response components, $\xi_x$ and $\xi_y$, respectively, have similar mode shapes for all ARs. The spanwise component has a double-peak structure becoming more apparent with increasing AR, which implies that streamwise vortex, and the lift-up mechanism, becomes more evident at higher AR. The double-peak in the spanwise component of the minimal-norm forcing, $\phi_z$ seen at every AR indicates that the optimal forcing for wall shear has always a vortex-like structure. We do not see such a structure in the correlated forcing, $\bm{\chi}$. Note that the correlated forcing contains both solenoidal and non-solenoidal parts, which may result in disappearance of the vortex shape. A similar observation was reported in \cite{morra_jfm_2021}. 

The similarity analysis provided in this subsection involves only the mode shapes ignoring any amplitude information. To investigate the trends in the mode amplitudes, we plot also the peak values for the absolute value of each mode component in figure \ref{fig:modeAmp}, this time without normalizing the modes to have unit wall shear. We see in figure \ref{fig:modeAmp} that mode amplitudes of correlated velocity decay with constant power with increasing $\beta^+$, with the decay rate being $\sim$2. The only exception is at AR=2, which is shown with red color. The decay rate is nearly the same for all three components at all ARs from 3 to 10.  Mode amplitude increases up to AR=6, and then saturates. In the minimal-norm forcing, $\bm{\phi}$, amplitude of the streamwise component is seen to be an order of magnitude smaller than the spanwise component. The amplitude difference increases with AR: $\phi_x$ becomes smaller while $\phi_z$ remains unchanged. Similar to the minimal-norm forcing, wall-normal and spanwise components of wall-shear-correlated forcing, $\chi_y$ and $\chi_z$, respectively, have nearly the same amplitudes with increasing AR. The amplitude of the streamwise component, $\chi_x$, on the other hand, increases up to AR=6 and saturates for higher AR, similar to the response mode.

\subsection{Self-similarity at sub-dominant wall-attached structures}

\begin{figure}
  \centerline{\resizebox{\textwidth}{!}{\includegraphics{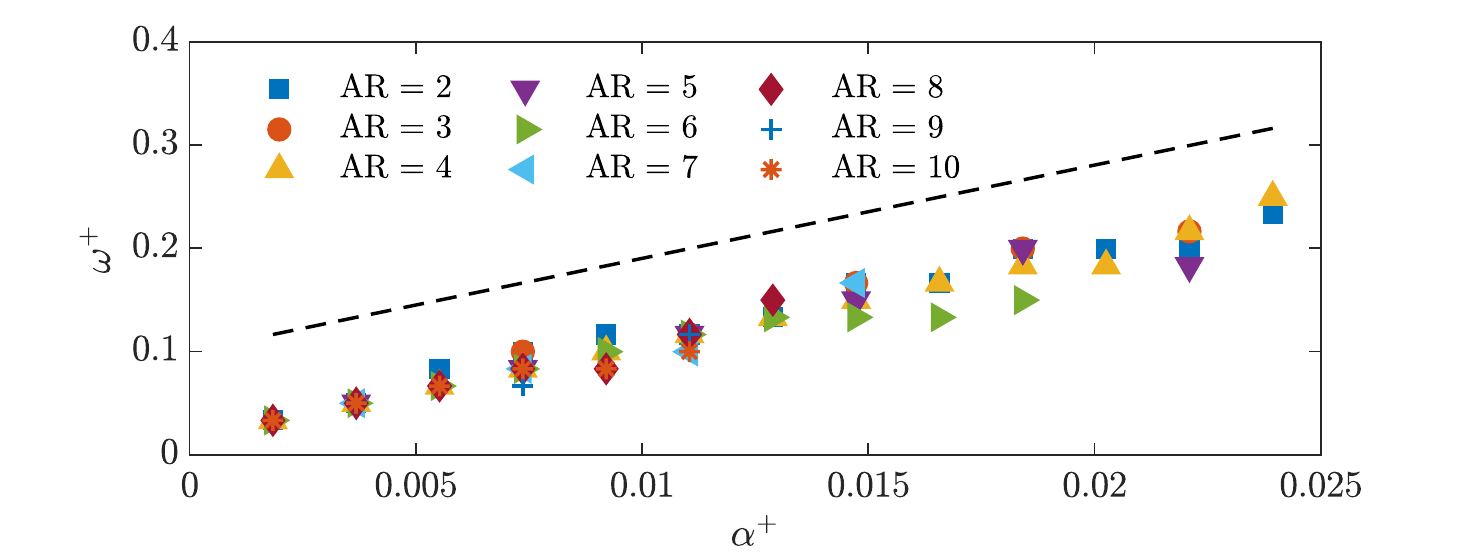}}}
  \caption{Shifted trend line for $\omega^+$ corresponding to sub-dominant wall-shear structures.}
\label{fig:omgmaxallshft}
\end{figure}

\begin{figure}
  \centerline{\resizebox{\textwidth}{!}{\includegraphics{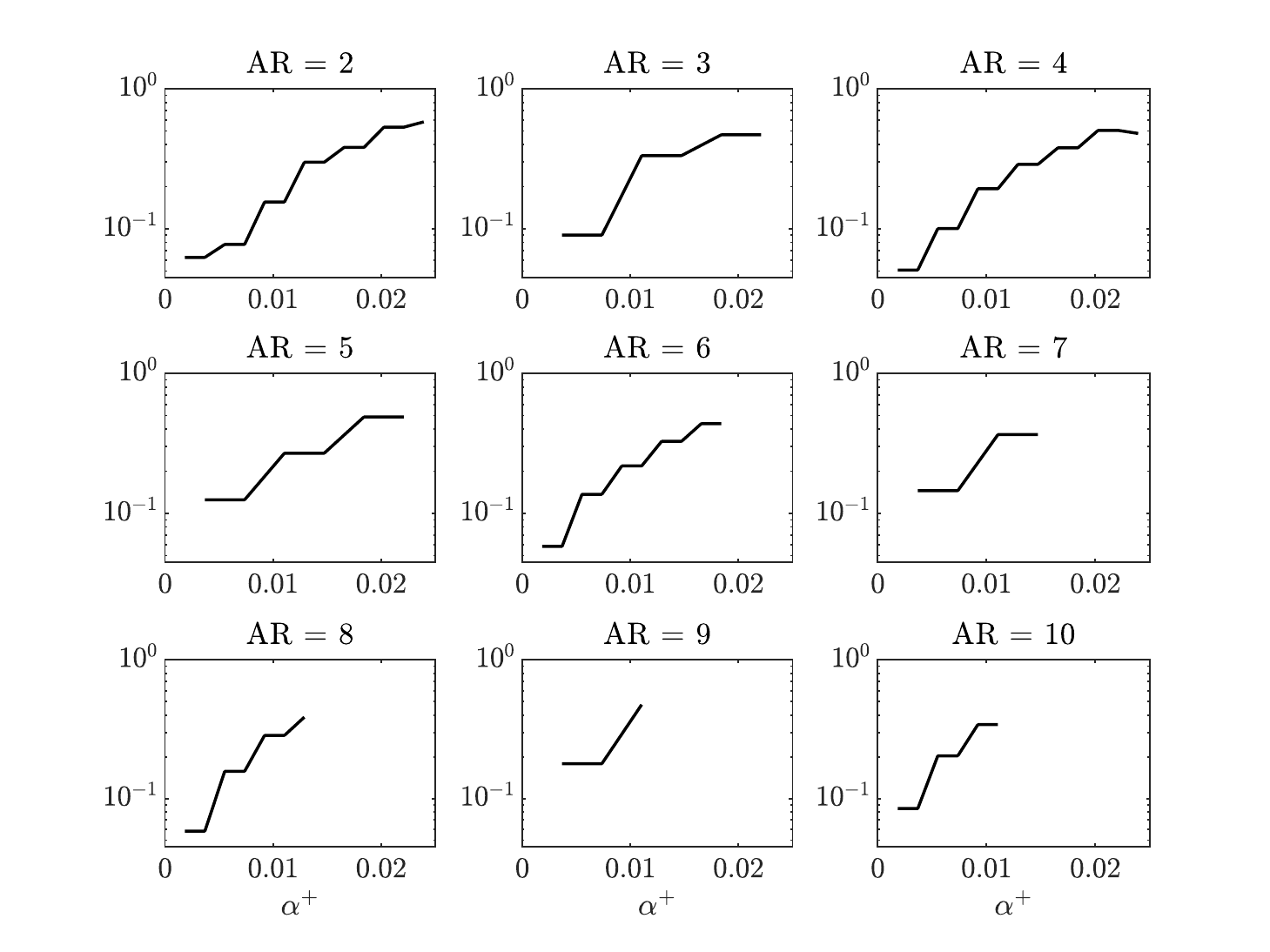}}}
  \caption{Ratio of the amplitude of $\tau_z$ calculated at sub-dominant frequencies to that calculated at $\omega_{max}$ at different wavenumbers.}
\label{fig:tauzratio}
\end{figure}

We will now investigate if the self-similarity observed in the dominant wall-attached structures extends to sub-dominant structures as well. To achieve this, we shift the trend line used to select the frequency for a given wavelength pair, as seen in figure \ref{fig:omgmaxallshft}. The ratio of the corresponding wall-shear spectra to that of the dominant structures investigated above is shown in figure \ref{fig:tauzratio}. We see that shifting to higher frequencies, the energy content at small wavenumbers drops significantly, while at high wavenumbers, it remains comparable to the energy in dominant structures. 

\begin{figure}
  \centerline{\resizebox{\textwidth}{!}{\includegraphics{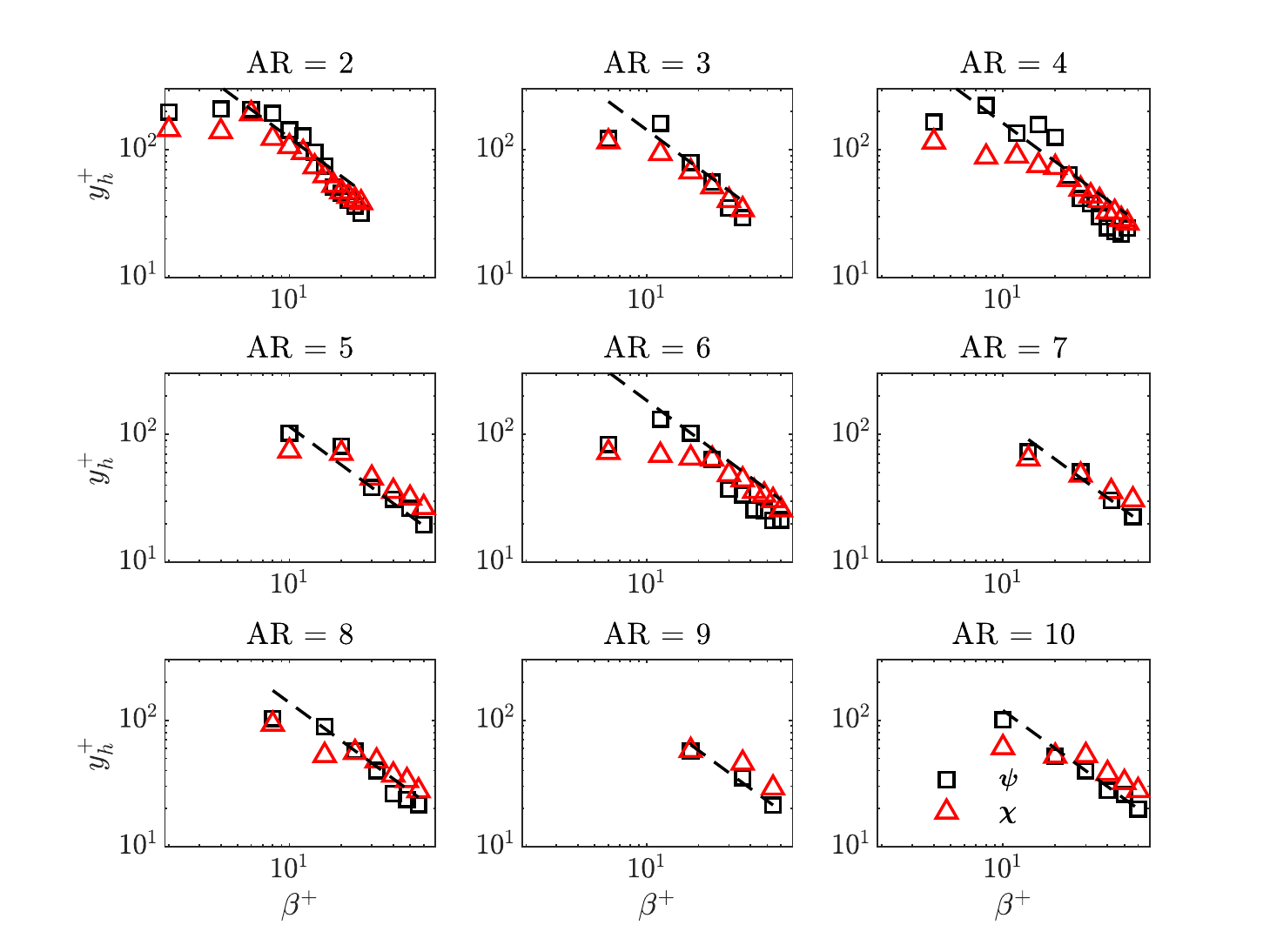}}}
  \caption{Mode center, $y_h^+$ obtained using $\xi_x$ (square) and and $\chi_z$ (triangle) at different wavenumbers at sub-dominant frequencies. Each subplot shows results for a different AR. The dashed lines show $1/{\beta^+}$ trend.}
\label{fig:yhvsbetaallshft}
\end{figure}

\begin{figure}
  \centerline{\resizebox{\textwidth}{!}{\includegraphics{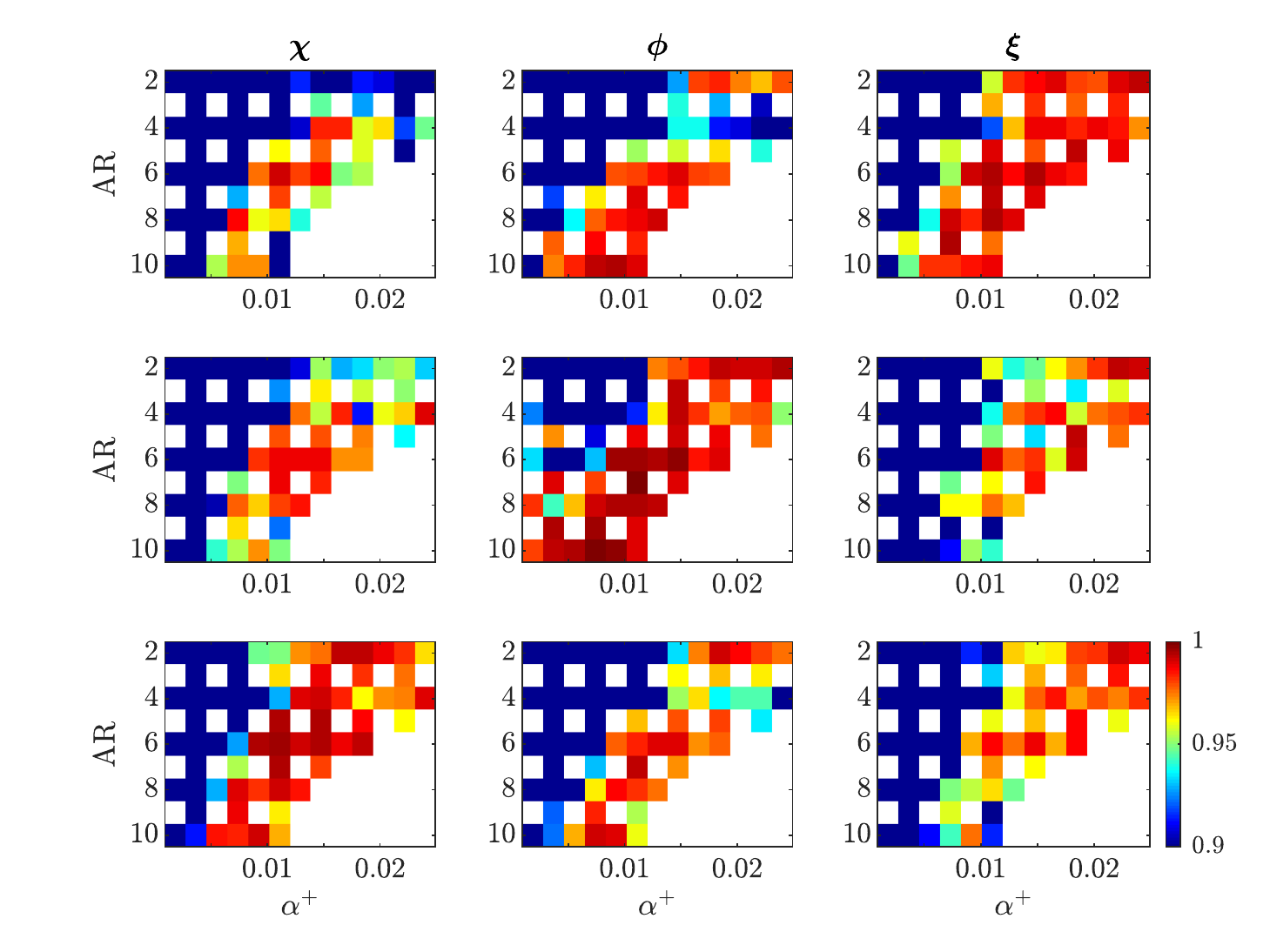}}}
  \caption{Self-similarity maps, $\gamma_{\bm{\chi}}$ (left), $\gamma_{\bm{\phi}}$ (center), and $\gamma_{\bm{\xi}}$ (right) for different ARs and $\alpha^+$ values obtained at sub-dominant frequencies. Rows correspond to the $x$-, $y$- and $z$-components from top to bottom.}
\label{fig:selfsimmapshft}
\end{figure}

We calculate mode center, $y_h$ based on $\xi_x$ for these sub-dominant structures, and investigate its change with $\beta$ in figure \ref{fig:yhvsbetaallshft}. It is observed that $\beta^{-1}$ trend appears only at wavenumbers where $\tau_z$ contains energy at levels comparable to the dominant structures. Figure \ref{fig:selfsimmapshft} shows the self-similarity maps, $\gamma_{\bm{\chi}}$, $\gamma_{\bm{\phi}}$, and $\gamma_{\bm{\xi}}$ for these sub-dominant structures similar to figure \ref{fig:selfsimmap}. Consistent with the $y_h$ vs. $\beta$ trends in figure \ref{fig:yhvsbetaallshft}, we observe self-similarity only at wavenumbers that contain high energy. The analyses was repeated for different shift values for $\omega$ where similar results were obtained but not presented here for brevity. These results imply that high-energy containing wall-attached structures, even if sub-dominant, and the associated forcing show self-similarity at a wide range of wavenumbers and ARs. Parallel to the lack of self-similarity in the wall-normal component of the forcing, the reason for not observing self-similarity in the low-energy modes may be once again slower convergence rates for these modes.

\section{Conclusions} \label{sec:conc}
We investigated the self-similarity of wall-attached structures in turbulent channel flows with a view to identifying attached eddies in the boundary layer. A flow database, obtained by direct numerical simulation (DNS), is considered with Reynolds number $Re_\tau=543$. The analysis is based on the attached-eddy hypothesis (AEH) \citep{townsend_jfm_1976,perry_jfm_1982}, which holds that attached eddies are high-energy containing structures that dominate the logarithmic region of the boundary layer, and whose size is determined by their distance to the wall. Guided by this definition, we explore self-similarity by considering the wall-normal organisation of streamwise-spanwise Fourier-mode pairs associated with a fixed aspect ratio; we then identify flow structures and forcing modes associated with wall shear in the spanwise direction, $\tau_z$. Spanwise wall shear was chosen as the representative quantity for quasi-streamwise vortices. The single-point measurement considered here amounts to a rank-1 system, which allows, within the linear framework provided by the resolvent analysis \citep{hwang_jfm_2010,mckeon_jfm_2010,cavalieri_amr_2019,
lesshafft_prf_2019}, identification of the flow structure associated with the wall-shear, i.e., wall-attached structures, and the associated forcing. We used resolvent-based spectral proper orthogonal decomposition (RESPOD) to perform an identification of wall-attached response and forcing modes. The method was compared to the resolvent-based-estimation (RBE) approach of  \citet{towne_jfm_2020}, which yields the minimal-norm forcing to generate a given response. {We observe that, using a linear scaling between the frequency and the streamwise wavenumber, the resulting wall-attached structures exhibit self-similarity in line with the AEH for a wide range of wavenumbers and aspect ratios. The frequency scaling is determined regarding the linear trend observed in the peak frequency of the power spectral density (PSD) of $\tau_z$ with respect to the streamwise wavenumber. Keeping the slope of the linear scaling constant, we observe that self-similarity extends to structures at sub-dominant frequencies as well. Although we do not particularly seek for energetic structures, it was shown in \cite{cheng_jfm_2020}, where a number of turbulent channels including one at Re=550 was investigated, that the wall-attached structures are dominant regarding the energy they contain. As we observe self-similarity in wall-attached structures in a wide range of frequencies and aspect ratios, we expect these structures to be relevant in terms of the AEH.} 

We extend the analysis to investigate such self-similarity for the forcing structures, again, associated with the wall-shear. Our findings reveal that both the minimal-norm forcing required to obtain the measured wall-shear dynamics (obtained by RBE) and the wall-shear-correlated forcing (obtained by RESPOD) that also drives the wall-attached flow structures exhibit self-similar behaviour.

The self-similarity of forcing modes obtained using RBE can be associated with the works of \citet{hwang_jfm_2010}, \cite{moarref_jfm_2013} and {\citet{sharma_ptrsa_2017}} among others, where it was shown that self-similarity is a property of the resolvent operator. Given the SPOD modes that are self-similar, it may be expected that associated forcing modes predicted by RBE would be self-similar. But the demonstration of a forcing self similarity using the RESPOD approach illustrates how that self similarity is indeed present in the actual forcing data, suggesting a self-similar organisation of the non-linear scale interactions that drive attached eddies. The self-similar forcing structures are shown to lead to elongated streaky structures, that are sustained by the lift-up mechanism, consistent with the discussion of \citet{cossu_rsta_2017} in which streaks and streamwise vortices participate in the lift-up mechanism as two interconnected elements of a single attached eddy.

The self-similar forcing educed from the DNS data may be used to construct dynamical models of attached eddies, following ideas similar to the works of \cite{moarref_jfm_2013}, \cite{hwang_jfm_2020} and \cite{skouloudis_prf_2021}. These works superpose coherent structures obtained with the linearised Navier-Stokes operator such that the the overall Reynolds stresses are obtained. 
As presented in this work, RESPOD is an appropriate technique to obtain the forcing that is coherent with a given flow response. The observed self-similarity of the forcing modes may be included in dynamic models of attached eddies for improved predictions of flow properties. 

Another promising direction is the use of the identified self-similar forcing to build flow estimators from a limited number of sensors. Linear estimators require assumptions of the forcing, and the quality of predictions depends on the accuracy of the forcing statistics included in the estimation \citep{chevalier_jfm_2006,martini_jfm_2020,amaral_jfm_2021}. For wall turbulence, the specification of a self-similar forcing may be a viable approach to construct accurate estimators, especially for high Reynolds numbers for which one cannot obtain DNS data.

Studies of coherent structures in flows have greatly benefited from the analysis of the linearised Navier-Stokes operator, which, by its non-normality, leads to more amplified structures that are more likely observed in a turbulent flow. However, a more refined view is obtained if the actual non-linear terms, which constitute a forcing in resolvent analysis, are used in the input. The framework developed in this work is not restricted to wall turbulence, and may be used in other flows to extract the properties of non-linearities driving the observed coherent structures.\\

\textbf{Funding.} This work has received funding from the Clean Sky 2 Joint Undertaking under the European Union’s Horizon 2020 research and innovation programme under grant agreement No 785303. U.K. has received funding from TUBITAK 2236 Co-funded Brain Circulation Scheme 2 (Project No: 121C061).\\

\textbf{Declaration of Interests.} The authors report no conflict of interest.

\appendix


\section{Linear operators used in resolvent analysis of channel flow} \label{app:1}
The linear operators used in \eqref{eq:nsres} are given as,
\begin{align}
\mathbf{A}=\begin{bmatrix}
\mathcal{L}_k & -\nabla \\
\nabla^T & 0
\end{bmatrix},\; 
\mathbf{B} =\mathbf{M}= \begin{bmatrix}
\mathbf{I} & 0 & 0 \\
0 & \mathbf{I} & 0 \\
0 & 0 & \mathbf{I} \\
0 & 0 & 0
\end{bmatrix}, \textrm{ and }
\mathbf{C}=\begin{bmatrix}
\mathbf{I} & 0 & 0 & 0\\
0 & \mathbf{I} & 0 & 0\\
0 & 0 & \mathbf{I} & 0
\end{bmatrix}
\end{align}
where $\mathcal{L}_k$ is the spatial linear N-S operator \citep{mckeon_jfm_2017} and given as,
\begin{align}
\mathcal{L}_k=\begin{bmatrix}
-i\alpha U_x + \nabla^2/Re & -\partial_y U_x & 0 \\
0 & -i\alpha U_x + \nabla^2/Re & 0 \\
0 & 0 & -i\alpha U_x + \nabla^2/Re
\end{bmatrix},
\end{align}
where $\nabla=[i\alpha,\,\partial_y,\,i\beta]^T$ and $\nabla^2=\partial_y^2 - (\alpha^2 + \beta^2)$ are the gradient and Laplace operators, respectively.

\FloatBarrier

\bibliographystyle{jfm}
\bibliography{biblio}

\end{document}